\documentstyle[preprint,prb,epsf,aps]{revtex}
\tighten
\begin{document}
\draft
\title{Coherent Ultrafast Optical Dynamics 
of the Fermi Edge Singularity
}
\author{N. Primozich, T. V. Shahbazyan, 
and I. E. Perakis}
\address{Department of Physics and Astronomy,  
Vanderbilt University, Nashville, TN 37235}
\author{D. S. Chemla}
\address{Department of Physics,  
University of California, Berkeley, CA 94720
and Materials Science Division, Lawrence Berkeley National
Laboratory,  Berkeley, CA 94720
}

\maketitle
\begin{abstract}
We develop a non--equilibrium many--body theory of the 
coherent femtosecond  nonlinear optical response of the 
Fermi edge singularity. We study the role of the 
dynamical Fermi sea response in the time--evolution of 
the pump--probe spectra. The electron--hole correlations 
are treated nonperturbatively with the time--dependent
coupled cluster expansion combined with the effective 
Hamiltonian approach. For short pulse durations, we find 
a non--exponential decay of the differential transmission 
during negative time delays, which is  governed by the 
interactions. This is in  contrast to the results obtained 
within the Hartree--Fock approximation, which predicts 
an exponential decay  governed by the dephasing time. 
We discuss the role of the optically--induced dephasing 
effects in the coherent regime.
\end{abstract}

\narrowtext

\section{\bf INTRODUCTION}
Ultrafast nonlinear spectroscopy provides
unique and powerful tools for studying
the dynamics of Coulomb correlation effects
in semiconductors, since the time resolution
can now be much smaller than 
the scattering times of the elementary excitations
and, 
e.g, the period of optical phonons or plasmons.
During the past decade, femtosecond pump--probe
and wave--mixing experiments have demonstrated
that in these materials both coherent and dissipative
processes are governed by many--body effects.\cite{nnlo} 
Parallel theoretical advances have
accounted for the two--particle correlation
effects in undoped semiconductors through the  time
dependent  Hartree--Fock approximation (HFA),
which has developed into  the very successful
Semiconductor Bloch Equations (SBE's)
formalism.\cite{stef,binder,koch} 
However, more recent experimental
observations could not be explained within this 
mean field approach and were attributed to four--particle 
and higher order Coulomb correlation 
effects.\cite{kne98,bartels99,schulzgen99,aoki99} These
high order correlation effects have been mostly
studied with equation--of--motion  methods for
density matrices \cite{che98,muk,axt}
or  pair--operators\cite{ost95} and  by using the 
Keldysh Green's function formalism.\cite{haug} 

At the same time, much less is known about the role
of {\em coherent} many--body effects in the nonlinear
optical response of doped semiconductors and, in
particular, modulation--doped heterostructures. Some
aspects of the effects of the electron--electron ({\em e--e})
scattering  have been 
described within the dephasing and relaxation time
approximations without microscopic  justification. Most
importantly however,  the role of the time--dependent
electron--hole ({\em e--h}) correlations between the
photoexcited holes and the Fermi sea (FS) electrons
on the ultrafast dynamics
is still poorly understood. The main difficulty
comes from the non--perturbative nature of the
Fermi edge singularity (FES) which dominates the
absorption spectrum close to the absorption onset,
and which cannot be described within the above
approximations.\cite{rev1,andrei} Important here
is that the {\em e--h } interactions between a  heavy 
photoexcited hole and the FS-electrons lead to the
scattering of a macroscopic number of low--energy
FS-pairs, which readjusts  the entire FS into
a new orthogonal scattering  state
in the course of the optical excitation 
(Anderson orthogonality catastrophe\cite{ander}).

The  FES is a   many--body resonance that has been
observed both in doped semiconductors and in
metals.\cite{rev1,fes,xray} Its non--Lorentzian lineshape can be
viewed as originating from the decay of an excitonic bound state
caused by its interactions with the gapless FS 
excitations.\cite{andrei} In a first approximation,
the excitonic effects in a doped semiconductor can
be described by extending the mean field approach (HFA)
to include  the effects of the screening and the Pauli
blocking due to the FS.\cite{rev1,andrei} However,
such a static treatment of the FS leads to 
a spurious bound state with respect to the Fermi energy,
$E_{F}$, referred to in the following as 
the HFA bound state.\cite{mahan}
This  discrete state, with binding energy $E_M$,
would appear at the energy $E_{F} - E_M$. Obviously,
for $E_M < E_{F}$ (as in typical modulation--doped quantum wells), 
such a state cannot exist since it overlaps with the FS continuum,
with which it interacts via the {\em e--h} potential.\cite{rev1,andrei}
The ``unbinding'' of this HFA bound state occurs via its
interactions  with the FS excitations,\cite{andrei}
 which are  {\em not} taken into account in the HFA.
Note that, in two--dimensional systems, a static FS
allows for bound states even for arbitrarily small
attractive interactions and therefore such an unbinding
cannot arise from static screening or from Pauli blocking effects. 
This spurious bound state could be artificially merged
with the continuum by introducing a dephasing time
comparable to its binding energy. Such an approximation,
however, neglects competely the interactions between
the photoexcited {\em e--h} pair and the FS excitations.
A microscopic description of the unbinding of the HFA bound state
presents a nontrivial problem due to the
non--perturbative nature of the {\em e--h} correlations
between the photoexcited hole and the FS excitations.\cite{andrei} 
Even within Green function techniques, this problem is rather
involved because vertex correction diagrams with arbitrarily
many crossed {\em e--h} interaction lines are as
divergent as the ladder diagrams and should be
treated on equal footing.\cite{mahbook} To perform
such a task, one must sum up at least the parquet
diagrams and  address the three--body correlations
between the photoexcited hole and a FS
excitation.\cite{gavor,andrei} 
Therefore, alternative methods
were developed for the case of linear absorption,
based on Fermi's golden rule with many--electron
eigenstates expressed in terms of Slater determinants.\cite{xray}
Such approaches become exact in the limit 
$m_{e}/m_{h} \ll 1$ ($m_{e}$ and $m_{h}$ are the
electron and hole masses respectively) and describe
quite accurately the FES lineshapes observed in
typical modulation--doped quantum wells.\cite{xray,ohtaka,sham1,hawr2}

Another approach to the FES problem is based on the
coupled cluster expansion (CCE).\cite{ccm,schon} This
general many--body technique provided exact results
in the limit
$m_{e}/m_{h} \ll 1$  \cite{ilias4} 
and was used to treat the hole recoil correlations in one
dimension.\cite{ilias5} In the latter case, an exact
solution was obtained for $m_{e}=m_{h}$. The CCE
has also been used to describe the
{\em e--e} correlation effects \cite{corr}. More
importantly, however, this method is well--suited for
describing correlations in non--equilibrium systems,
where it retains the advantages of diagrammatic
expansions without resorting to the Kadanoff-Baym
ansatz or to the Markovian approximation.\cite{neg} 

Let us now briefly summarize recent experimental 
studies of dephasing and relaxation processes in
modulation--doped quantum wells. Wang {\em et al.}
\cite{wang} performed pump--probe measurements under
resonant excitation conditions and observed an
incoherent redshift of the FES at positive time delays 
due to the incoherent bandgap renormalization. At
room temperature, Knox {\em et. al.} \cite{knox} 
measured very fast e--e scattering times estimated
$\sim $ 10 fs, which implies extremely fast dephasing
and relaxation.
At low temperatures, Kim {\em et al.} \cite{kim}
measured long dephasing times within the frequency
range $\omega < E_{F}$ (the frequency domain of the
FES), consistent with measurements in metals
\cite{metals}. This result is consistent with the
Landau Fermi liquid theory \cite{pines} (which
applies both to metals and doped--quantum wells
due to the similar values of  $r_{s}$) and points
out that, unlike for the {\em e--h} correlations, 
the {\em e--e} scattering is suppressed within the
frequency range of the FES. Furthermore, the above
experiment suggests that the hole dephasing times
are rather long. 
The  above experimental results were interpreted 
within a two--level system approximation (which
neglects all FES  correlation effects) by introducing
energy--dependent dephasing times obtained from the
quasi--equilibrium {\em e--e} self--energies.\cite{hawr1} 
Note here that, unlike in undoped semiconductors, the
relaxation of {\em real} photo-excited {\em e--h} pairs
due to their interactions with the FS
electrons\cite{metals} can obscure the role of
the {\em e--h} correlations.
However, these incoherent effects
can be suppressed by tuning the pump frequency 
{\em below} the FES resonance and thus exciting only
virtual carriers.
In this case,  coherent effects will dominate. 
Brener {\em et. al.}\cite{igal,ilias6} 
performed such pump--probe measurements
on doped QWs with well--below--resonance pump excitation 
and observed qualitatively different nonlinear
optical dynamics for the FES
as compared to excitons
in undoped QWs. This suggests that the differences in
the nature of the two resonances do manifest themselves
in the transient changes of the optical properties. 

In this paper, we develop a microscopic many--body 
theory for treating the coherent nonlinear optical
response of the FES. In particular, we focus on the
manifestations of the non--perturbative {\em e--h}
correlations 
(dynamical FS response) on the ultrafast
time evolution of the pump--probe spectrum. In order
to describe the dynamical FS response, we use a
general formalism developed recently \cite{per99}
(which we summarize in Appendix A).
Within this approach, the transient pump--probe
nonlinear absorption spectrum is mapped onto the
linear absorption of a ``pump--dressed'' system,
described by a time--dependent effective Hamiltonian.
This allows us to treat the non--equilibrium {\em
e--h} correlation effects non--perturbatively
with  the
time--dependent version of the CCE. We mostly focus on
negative time delays and off--resonant excitation
conditions, in which case the coherent effects dominate. 

In doped semiconductors, the pair--pair interactions,
analogous to the exciton--exciton interactions
in undoped semiconductors, are screened out,
while the {\em e--e} scattering is suppressed 
for below--resonant excitation as dictated 
by Fermi Liquid theory. Instead,
it is the   {\em e--h} interactions 
between the photoexcited {\em e--h} pairs and the FS carriers
that strongly affect the coherent ultrafast dynamics of the FES. 
An important difference from undoped semiconductors 
is that, due to its {\em gapless} pair excitation spectrum,
a FS responds {\em unadiabatically} 
to time--dependent perturbations.
In contrast, 
because of its finite Coulomb binding energy, an
exciton can be polarized by the pump optical field
without being ionized.
A non--equilibrium treatment beyond the HFA 
is necessary in order to take into account the unadiabatic 
time--dependent {\em change} in the {\em e--h} pair--FS 
interactions and the {\em e--h} scattering processes
induced by the ultrafast pump excitation. 
We study in detail the
qualitative differences in the ultrafast dynamics
with and without (HFA) the dynamical FS response
and point out the
important role of the {\em e--h} correlations in
the coherent regime.

The outline of this paper is as follows. In
Section II we summarize our formalism for the
nonlinear optical response and in Section III
we outline the basics of the CCE approach to
correlations. In Section IV we obtain the 
time--dependent parameters of the effective 
semiconductor Hamiltonian and the effective
optical transition matrix elements. In Section V
we describe the effects of the optical excitation 
and the many--body correlations on the dynamics
of the photoexcited {\em e--h} pair that determines the FES lineshape. 
In Section VI we present our results for the
nonlinear absorption of the FES and its
time--evolution, and discuss the role of the  
dynamical FS response. Section VII concludes
the paper. 


%
\section{TRANSIENT NONLINEAR ABSORPTION SPECTRUM}

We start by recapping the main points of our 
formalism\cite{per99} (see Appendix A).  
In the rotating frame,\cite{eber} the Hamiltonian 
describing this system is 
\begin{equation} \label{H-tot}
H_{{\rm tot}}(t) = H + H_{p}(t) + H_{s}(t) \, .
\end{equation}
The first term is the ``bare'' Hamiltonian,
\begin{equation} \label{H0} 
H =  \sum_{\bf {\bf q}}\varepsilon_{{\bf {\bf q}}}^{v}
b_{-{\bf {\bf q}}}^{\dag} b_{-{\bf {\bf q}}} + 
\sum_{{\bf {\bf q}}} (\varepsilon_{{\bf {\bf q}}}^{c}
+\Omega) a^{\dag}_{{\bf {\bf q}}}a_{{\bf {\bf q}}}
+V_{ee} +V_{hh}+V_{eh},
\end{equation}
where  $a^{\dag}_{{\bf {\bf q}}}$ is the creation
operator of a conduction electron with energy
$\varepsilon_{{\bf {\bf q}}}^{c}$ and mass $m_{e}$,
$b^{\dag}_{-{\bf {\bf q}}}$ is the creation operator
of a valence hole with energy
$\varepsilon_{{\bf {\bf q}}}^{v}$ and mass $m_{h}$,  
$V_{ee}, V_{eh}$, and $V_{hh}$ describe the
{\em e--e}, {\em e--h}, and {\em h--h} interactions,
respectively, and
$\Omega= E_{g} + E_{F}(1 + m_{e}/m_{h}) - \omega_{p} $
is the detuning of the central frequency of the
optical fields, $\omega_p$, from the Fermi level, $E_{g}$ being
the bandgap (we set $\hbar=1$). The second and third
terms describe the coupling of  the pump optical field,
${\cal E}_p(t) e^{i {\bf k}_p \cdot {\bf r}  - i\omega_{p} t}$,
and  the probe optical field,
${\cal E}_s(t)e^{i{\bf k}_s \cdot {\bf r} - i\omega_{p} (t-\tau)}$, 
respectively: 
\begin{eqnarray}\label{Hp}
H_{p}(t) &=& \mu {\cal E}_p(t) \left[ e^{i {\bf k}_{p} \cdot
 {\bf r }}U^{\dag}+ {\rm H.c.} \right], 
\nonumber\\
H_s(t) &=& \mu {\cal E}_s(t) \left[ e^{i {\bf k}_{s} \cdot
 {\bf r } + i \omega_{p} \tau}U^{\dag}+ {\rm H.c.} \right],
\end{eqnarray} 
where the pump amplitude ${\cal E}_p(t)$ is
centered at time $t=0$ and the probe amplitude
${\cal E}_s (t)$ is centered at the time
delay $t=\tau$,  $\mu$ is the interband
transition matrix element, and
\begin{equation}
U^{\dag} = \sum_{{\bf {\bf q}}}
a^{\dag}_{{\bf {\bf q}}}
b^{\dag}_{-{\bf {\bf q}}} \, 
\label{trans-op}
\end{equation} 
is the optical transition operator. In Appendix B we clarify our
convention for the time delay $\tau$ and relate
it to the most commonly used conventions
in pump/probe and four wave mixing (FWM) experiments.

In many experiments, the amplitude of the probe field is much
smaller than that of the pump, 
$|{\cal E}_p(t)| \gg |{\cal E}_s (t)|$. 
In that case, as was shown in Ref.\ \onlinecite{per99} 
(see Appendix A),
the experimentally--measurable nonlinear optical
polarization can be obtained in terms of the
linear response functions of a ``pump--dressed''
semiconductor to a probe field
(within $\chi^{(3)}$, this is true even for comparable 
pulse amplitudes). This ``dressed'' system is 
described by a time--dependent effective Hamiltonian
$\tilde{H}(t)$, which is obtained by performing a
time--dependent Schrieffer--Wolff/Van Vleck \cite{s-w,cohen}
canonical transformation on the Hamiltonian $H+H_{p}(t)$.
As we shall see in the next Section, in all the cases
of interest, the effective Hamiltonian $ \tilde{H}(t)$
has the same operator form as the bare Hamiltonian 
$H$, with the important difference that the band
dispersions (effective masses) and interaction
potentials are {\em time--dependent} through
${\cal E}_{p}(t)$. Thus, the calculation of the nonlinear 
absorption spectrum reduces to that of the linear 
absorption spectrum of the ``pump--dressed'' semiconductor 
with uncoupled ``effective bands'' --- a great
simplification that allows us to use 
straightforward generalizations of 
well established theoretical tools.
It is important to note that, as was shown in Ref. \onlinecite{per99} 
(see also Appendix A), 
such a ``dressed semiconductor'' 
approach is not restricted to monochromatic pulses 
and is valid for {\em any}  pulse duration. 
In Appendix A, we show that the pump--probe
nonlinear polarization has the following form : 
\begin{equation} 
P_{{\bf k}_{s}}(t) =
-i\mu^2   e^{i {\bf k}_{s} \cdot {\bf r}- i\omega_{p} (t - \tau)}
\int_{-\infty}^{t} dt' {\cal E}_{s}(t')\langle \Phi_0(t)| 
\tilde{U}(t){\cal K}(t,t')\tilde{U}^{\dag}(t')
|\Phi_0(t') \rangle \, .
\label{spec1}
\end{equation}
Here, $|\Phi_0(t)\rangle$ is the state evolved with 
$\tilde{H}(t)$ from the semiconductor ground state 
$| 0 \rangle$ of $H$, $\tilde{U}^{\dag}(t )$
is the effective optical transition operator
describing the  probability amplitude 
for the photoexcitation  of an {\em e--h} pair  
by the probe field in the presence of the pump
excitation, and ${\cal K}(t,t')$ is the 
time--evolution operator satisfying  the
Schr\"{o}dinger equation
\begin{equation}\label{eigen}
i\frac{ \partial}{\partial t}{\cal K} (t,t') 
=\tilde{H}(t){\cal K} (t,t') \, .
\end{equation}
The above equation  describes the time evolution of a 
probe--photoexcited {\em e--h} pair
in the presence of the pump excitation. The
effective Hamiltonian and effective transition
operator are given by (see Appendix A)
\begin{equation} 
\tilde{H}(t) = H_{0} - \frac{\mu}{2}\, 
\left({\cal E}_p(t)\left[\hat{\cal P}(t), U^{\dag} \right]
+ {\rm H.c.} \right),
\label{eff}
\end{equation} 
and 
\begin{equation} 
\tilde{U}^{\dag}(t) 
= U^{\dag} + \frac{1}{2} 
\left[ \hat{\cal P}(t), \left[U^{\dag}, \hat{\cal P}^{\dag}(t)\right]
\right] + \frac{1}{2} 
\left[\hat{\cal P}^{\dag}(t), \left[ U^{\dag}, \hat{\cal P}(t)\right]
\right], \label{me}
\end{equation} 
where the  operator 
$\hat{\cal P}^{\dag}(t)$, which generates the canonical 
transformation, satisfies the equation 
\begin{equation}
i   \frac{\partial
\hat{\cal P}^{\dag}(t)}{\partial t} =
\left[ H,\hat{\cal P}^{\dag}(t)
\right] - \mu {\cal E}_p(t) U^{\dag},
\label{sig}
\end{equation}
with the initial condition $
\hat{\cal P}^{\dag}=0$ before
the pump arrives.
Eqs. (\ref{eff}) and  (\ref{me}) include  all the
pump--induced corrections to $\tilde{U}^{\dag}(t)$ 
and $\tilde{H}(t)$ up to the second order in the pump
optical field and are valid when
$\left( d_{p}/\Omega \right)^{2} \lesssim 1$ (for
off--resonant excitation) or 
$\left( d_{p} t_{p}\right)^{2} \lesssim1$ (for
resonant excitation), where $t_{p}$ is the pump
duration.

It should be emphasized that, although
 Eq.\ (\ref{eff}) 
gives the effective Hamiltonian 
up to the  second order in the pump field,
${\cal E}_p{\cal E}_p^{\ast}$,  
the polarization 
expression Eq.\ (\ref{spec1}) describes the effects of 
$\tilde{H}$
in {\em all} orders.
For example, as we
shall see in Section IV, the pump--induced term in
$\tilde{H}$ contains  self--energy
corrections to the electron/hole energies, 
which describe 
(among other effects) the resonance blueshift due to
the ac--Stark effect. Although the {\em magnitude} of
these self--energy corrections, calculated 
using Eq.\ (\ref{eff}),  
is quadratic in
${\cal E}_p$, the
correct position of the resonance 
can only be obtained
by evaluating the pump/probe polarization (\ref{spec1})
nonperturbatively (beyond $\chi^{(3)}$), i.e., without
resorting to the expansion of the time--evolution
operator ${\cal K}(t,t')$ in the pump field. 
Importantly, the same is 
true when calculating the effects of the 
self--energy corrections 
on the {\em e--h} correlations. 
As we shall see in Section VI, such a nonperturbative (in
the pump field) treatment of the non--linear response of
the FES is crucial for the adequate description of
the pump/probe spectrum at negative time delays.
In Section V,
we will describe the corresponding procedure, which
accounts  for the FS dynamical response. 
In contrast, the third--order polarization, $\chi^{(3)}$,
can be simply obtained from Eq.\ (\ref{spec1}) by
expanding ${\cal K}(t,t')$ to the {\em first} order
in the pump--induced term in $\tilde{H}$ [second term
in Eq.\ (\ref{eff})]. We did not include in Eq.\ (\ref{spec1}) 
the  biexcitonic contribution  
(coming from the excitation of two {\em e--h} pairs by the pump 
{\em and} the probe pulses)
since it vanishes for the negative time delays ($\tau<0$)
of interest here.\cite{per99}

The important advantage of  Eq.\ (\ref{spec1}), as
compared to the  equations of motion for the
polarization, comes from its similarity to the
linear polarization that determines the linear
absorption spectrum \cite{mahbook}. This can be seen
by setting ${\cal E}_p(t) =0$ in the Eqs. (\ref{eff})
and  (\ref{me}), in which case the effective
time--evolution and optical transition operators
transform into their ``bare'' counterparts:
${\cal K}(t,t') \rightarrow  e^{-i  H (t - t' )}$ and
$\tilde{U}^{\dag}(t) \rightarrow U^{\dag}$. Moreover, like
$U^{\dag}$, the effective transition operator
$\tilde{U}^{\dag}(t)$ creates a single {\em e--h} pair,
while, as we shall see below, the effective Hamiltonian
$\tilde{H}(t)$ can be cast in a form similar to
$H$. This allows one to interpret the Fourier transform
of  Eq.\ (\ref{spec1}) as the linear absorption spectrum 
of a ``pump--dressed'' semiconductor with two uncoupled
 but time--dependent effective bands. This mapping  
simplifies significantly the  calculation of the FES 
ultrafast nonlinear optical response by allowing a
straightforward generalization of the CCE.


%
\section{COUPLED--CLUSTER EXPANSION}

In this Section, we show how the time--dependent 
CCE can be used to study the effects of the
{\em e--h} correlations (dynamical FS response) on the
time evolution of  the {\em e--h} pair photoexcited
by the probe. Our goal is to evaluate the many--body
state $| \Psi(t) \rangle = {\cal K}(t,t')
\tilde{U}^{\dag}(t') |\Phi_0(t') \rangle$ that enters
in  Eq.\ (\ref{spec1}). This state satisfies the  Schr\"{o}dinger
equation
\begin{equation}
i   \frac{ \partial}{\partial t} | \Psi(t) \rangle 
=\tilde{H}(t) | \Psi(t) \rangle.
\label{eigen1}
\end{equation}
As already mentioned, $\tilde{H}(t)$ has the same
form as the bare Hamiltonian $H$. This allows us to obtain
$| \Psi(t) \rangle$ through a straightforward
generalization of the linear absorption calculations
\cite{schon,ilias4,ilias5}. After eliminating the
valence hole degrees of freedom, \cite{llp,ilias5}
$| \Psi(t) \rangle$ is expressed in the CCE form

\begin{equation}\label{ccw}
| \Psi(t) \rangle =  e^{S(t)} |\Phi(t) \rangle,
\end{equation}
where  the time--dependent operator $S(t)$
creates FS--pairs and is given by 
\begin{equation} 
S(t) = \sum_{p > k_{F},  k <k_{F}} 
s({\bf p,k},t) a^{\dag}_{{\bf p}}   
a_{{\bf k}} + \sum_{p_{1}, p_{2} >k_{F},
k_{1}, k_{2}<k_{F} } s_{2}({\bf p_{1},p_{2},k_{1},k_{2}},t) 
a^{\dag}_{{\bf p_{1}}} a^{\dag}_{{\bf p_{2}}} a_{{\bf k_{2}}}
a_{{\bf k_{1}} }
+ \cdots, \label{S}
\end{equation}
while the  state $ |\Phi(t) \rangle$, discussed in
Section V, describes the time evolution of the
probe--induced {\em e--h} pair. In Eq.\ (\ref{S}),
the amplitude $s({\bf p}, {\bf k}, t) $ describes
the {\em e--h} correlations which, in particular,
are responsible for the unbinding of the HFA bound state; 
the two--pair amplitude $s_{2}$
describes the {\em e--e} interaction effects at the 
RPA level and beyond.
>From a physical point of view, the operator 
$e^{S(t)}$ describes the readjustment 
of the FS density profile 
during the optical transition 
in response to the FS  
interactions 
with the photoexcited {\em e--h} pair.

Substituting  Eq.\ (\ref{ccw}) into the Schr\"{o}dinger
equation Eq.\ (\ref{eigen1}), multiplying by the
operator $e^{-S(t)}$ from the lhs, and using the
fact that $[S(t), S(t')]=[\dot{S}(t), S(t')]=0$, we obtain
\begin{equation} \label{phi}
i\frac{\partial}{\partial t}|\Phi(t) \rangle + 
i\dot{S}(t)|\Phi(t) \rangle 
=  e^{-S(t)} \tilde{H}(t) e^{S(t)}| \Phi(t) \rangle,  
\end{equation}
where the transformed Hamiltonian on the rhs 
can be expressed in terms of the   commutator series
(Baker--Campbell--Hausdorff expansion)
\begin{equation} \label{trH}
e^{-S(t)}  \tilde{H}(t) e^{S(t)}
= \tilde{H}(t) 
+ [ \tilde{H}(t), S(t) ] 
+ \frac{1}{2} [[H, S(t)],S(t) ] + \cdots  
\end{equation}
An important advantage of the CCE is that, 
due to the FS momentum  restrictions in  Eq.\ (\ref{S}), 
the above series {\em terminates}
after the first few terms 
(three for quartic interactions) and a closed--form  expression of
the transformed Hamiltonian (\ref{trH}) can be obtained  in terms of
$S(t)$. By requiring that all FS--pair creation processes are
eliminated from the above equation, we obtain the CCE
equations\cite{ccm,schon} for $S(t)$. Before proceeding
with such a calculation however, we derive in the
next Section explicit expressions for $\tilde{H}(t)$
and $\tilde{U}(t)$.

%
\section{EFFECTIVE HAMILTONIAN AND OPTICAL TRANSITION 
OPERATOR}

In this Section, we derive  second--quantized 
expressions for the effective  Hamiltonian
$\tilde{H}(t)$ and  optical transition operator
$\tilde{U}(t)$.  We start with Eqs.\ (\ref{eff})
and\ (\ref{me}), which express $\tilde{H}(t)$
and $\tilde{U}(t)$ in terms of the canonical
transformation operator $\hat{\cal P}^{\dag}(t)$.
The latter is given by Eq.\ (\ref{sig}), which 
includes the effects of the Coulomb interactions
on the {\em pump} photoexcitation. It is important to realize that
the effect of  Coulomb {\em e--h} interactions on the pump and
the probe
photoexcitations is very different. 
For an adequate description
of the FES, the {\em e--h} interactions should be taken into account
{\em non--perturbatively} for the {\em probe}--photoexcited
pair. Indeed, the nonlinear  absorption spectrum 
at a given frequency $\omega$ close to the 
FES resonance is determined by the time--evolution
for {\em long} times (of the order of
the dephasing time $T_2$)
of an {\em e--h} 
pair photoexcited 
by the {\em probe} at 
the  energy
$\omega$.\cite{mahan2}
Since the characteristic 
``{\em e--h} interaction time''
$E_M^{-1}$ (inverse HFA bound state energy)
that determines  the nonexponential polarization decay 
is much shorter that the 
dephasing time,  
the long--time asymptotics of the response function 
(to the {\em probe}) depends non--perturbatively on the 
{\em e--h} interactions.  
In contrast,  a short {\em pump}
optical pulse excites a wavepacket 
of {\em continuum} 
{\em e--h} pair states
(unlike in the discrete exciton case)
with energy width $\sim t_{p}^{-1}$,
which thus evolves during timescales 
comparable to the pulse duration $t_{p}$.
Also, the 
corrections to the effective Hamiltonian 
are determined by the time 
evolution of the 
pump--induced carriers 
only up to times  $\sim t_{p}$
(see Eq.\ (\ref{eff})).
Therefore, if the 
``{\em e--h} interaction time'' 
$E_M^{-1}$ is larger than 
$t_{p}$, $t_p E_M\lesssim 1$, 
(i.e., if the pump pulse
frequency width exceeds $E_M$), 
the {\em e--h} interactions can be
treated perturbatively when describing the time--evolution of the 
{\em pump}--photoexcited pairs. 
This   can also be  shown explicitly for 
the third--order nonlinear polarization. 
In the general expression for  $\chi^{(3)}$, 
all contributions that 
depend on the pump are integrated over the
width of the pump pulse; therefore, any resonant enhancement
of $\chi^{(3)}$  that depends on the  pump frequency
will be broadened  out for sufficiently short pulses 
with frequency width that  exceeds $E_M$. 
In other words, when deriving the
pump--renormalized parameters, one can
treat Coulomb interactions perturbatively if the above condition is
fulfilled. 
In fact, the above 
situation  is somewhat similar to the calculation
of the linear absorption spectrum close to the indirect
transition threshold, 
where perturbation theory can be used.\cite{andrei,gavor}
Thus the
above consideration applies  even for long  pulse  durations
provided that  the detuning $\Omega$ exceeds $E_M$.
This is certainly the case for the off--resonant 
excitation conditions considered in Section\ VI. 
However, in order to obtain the full absorption spectrum,  
the time--evolution of the {\em probe}--photoexcited
pair with such effective Hamiltonian (with perturbatively calculated
time--dependent parameters) should be treated 
{\em non--perturbatively}, as described in the following Section.

We now proceed with the derivation of the effective Hamiltonian.
To lowest order in the interactions,
$\hat{\cal P}^{\dag}(t)$ can be presented as 
\begin{eqnarray} \label{op1}
\hat{\cal P}^{\dag}(t) 
=  &&\sum_{{\bf q}} {\cal P}_{eh}({\bf q},t)
a^{\dag}_{{\bf q}} b^{\dag}_{-{\bf q}}
+ \frac{1}{2} \sum_{{\bf p,p',k}} 
{\cal P}_{eh}^{e}({\bf p p';k;}t)b^{\dag}_{{\bf k-p-p'}}  
a^{\dag}_ {{\bf p}} a^{\dag}_ {{\bf p'}}a_ {{\bf k}} 
\nonumber\\&&
+\frac{1}{2} \sum_{{\bf p, p',k}} 
{\cal P}_{eh}^{h}({\bf p p';k};t) 
a^{\dag}_ {{\bf p + p' - k}} 
b^{\dag}_{-{\bf p}} b^{\dag}_{-{\bf p'}}
b_{-{\bf k}},
\end{eqnarray}
where ${\cal P}_{eh}$ is the  probability amplitude 
for excitation of an e--h pair with zero momentum 
satisfying 
\begin{eqnarray}
\label{dir} 
i\frac{\partial}{\partial t}{\cal P}_{eh}({\bf q},t) 
=\left( \Omega + \varepsilon_{\bf{q}}^{(c)} +
\varepsilon_{\bf{-q}}^{(v)} - i \Gamma \right) 
{\cal P}_{eh}({\bf q},t)- \mu {\cal E}_p(t)-
\sum_{{\bf q}'} v({\bf q}-{\bf q}'){\cal P}_{eh}({\bf q'},t),
\end{eqnarray} 
where  $\Gamma=T_2^{-1}$
describes  the dephasing due to processes not included in
the Hamiltonian $H$ (e.g., due to  phonons). 
In Eq.\ (\ref{op1}), 
\begin{eqnarray}\label{eeh}
{\cal P}_{eh}^{e}({\bf pp';k};t)  
= i &&\int_{-\infty}^{t}dt'
e^{-i(t - t')  \left( \Omega + \varepsilon_{{\bf p}}^{c} 
+ \varepsilon_{{\bf p'}}^{c} 
- \varepsilon_{{\bf k}}^{c}
+ \varepsilon_{{\bf p+p'-k}}^{v} - i \Gamma \right)} \times 
\nonumber \\&&
\left\{ 
 v({\bf p}-{\bf k})
\left[{\cal P}_{eh}({\bf p'},t') -
{\cal P}_{eh}({\bf p+p'-k},t') \right] 
- ({\bf p} \leftrightarrow {\bf p'})
\right\}
\end{eqnarray} 
describes the scattering of the photoexcited
{\em e--h} pair with an electron, and 
\begin{eqnarray} \label{ehh}
{\cal P}_{eh}^{h}({\bf pp';k};t)  
= -i &&\int_{-\infty}^{t}dt'e^{-i(t-t') 
\left( \Omega + \varepsilon_{{\bf p+p'-k}}^{c} 
+ \varepsilon_{{\bf p'}}^{v} + \varepsilon_{{\bf p}}^{v} 
- \varepsilon_{{\bf k}}^{v} - i \Gamma\right)} \times
\nonumber \\&&
\left\{ 
 v({\bf p}-{\bf k})
\left[{\cal P}_{eh}({\bf p'},t') -
{\cal P}_{eh}({\bf p+p'-k},t') \right] 
- ({\bf p} \leftrightarrow {\bf p'})
\right\}
\end{eqnarray} 
describes the scattering of the photoexcited
{\em e--h} pair with a hole. 
The above expressions 
describe in the lowest order in the screened
interaction\cite{ander1}
$\upsilon({\bf p} - {\bf k}) $ 
the coherent pump--induced processes, 
the effects of the Hartree--Fock 
pair--pair and pair--FS interactions, and the  dynamical FS
response  to the {\em  pump} photoexcitation.

By substituting Eq.\ (\ref{op1}) into Eq.\
(\ref{me}), we obtain the following expression
for the effective optical transition operator: 
\begin{eqnarray} \label{tme}
\tilde{U}^{\dag}(t) |\Phi_0(t) \rangle = 
\sum_{p > k_{F}} M_{{\bf p}}(t)  a^{\dag}_{{\bf p}} \,
b^{\dag}_{-{\bf p}} |0 \rangle + 
\frac{1}{4} \sum_{p,p'>k_{F},k<k_{F}}
M_{{\bf pp'k}}(t) a^{\dag}_{{\bf p}} \, a^{\dag}_{{\bf p'}} \, 
b^{\dag}_{{\bf k - p - p'}} \, a_{{\bf k}}  |0\rangle,
\end{eqnarray} 
where the effective matrix element $M_{{\bf p}}(t)$ 
includes corrections due to phase space filling and 
Hartree--Fock  interactions, and $M_{{\bf pp'k}}(t)$
is the probability amplitude for indirect optical
transitions \cite{andrei} induced by the pump optical
field, which contribute to the pump--probe
polarization in the second order in the
interactions. The explicit expressions for
$M_{{\bf p}}(t)$ and $M_{{\bf pp'k}}(t)$ are
given in Appendix C. 

We turn now to the effective Hamiltonian
$\tilde{H}(t)$. After substituting Eq.\
(\ref{op1}) into Eq.\ (\ref{eff}) 
we obtain that 
\begin{eqnarray} \label{Heff}
\tilde{H}(t)= 
\sum_{\bf q} \varepsilon^{v}_{{\bf q}}(t) 
b_{-{\bf q}}^{\dag}b_{-{\bf q}}
+\sum_{{\bf q}} \varepsilon^{c}_{{\bf q}}(t)
a^{\dag}_{{\bf q}}a_{{\bf q}} 
+ V_{eh}(t) + V_{ee}(t),
\end{eqnarray} 
where 
\begin{eqnarray}  \label{ed} 
\varepsilon^{c}_{{\bf q}}(t)=
\varepsilon_{{\bf q}}^{c} + \mu {\cal E}_p(t)
{\rm Re}\left[ {\cal P}_{eh}({\bf q},t) - 
\sum_{{\bf q}'} {\cal P}_{eh}^{e}({\bf qq';q};t)\right]
\end{eqnarray} 
is the effective conduction electron energy; 
\begin{eqnarray} \label{hd}
\varepsilon^{v}_{{\bf q}}(t)= E_{g} 
+ \varepsilon^{v}_{{\bf q}} + \mu {\cal E}_p(t) 
{\rm Re}\left[ {\cal P}_{eh}({\bf q},t) - 
\sum_{{\bf q}'} {\cal P}_{eh}^{h}({\bf qq';q};t)\right],
\end{eqnarray} 
is the effective  valence hole energy; 
\begin{eqnarray}
V_{eh}(t)= 
-\sum_{{\bf kk'q} }
\upsilon_{eh}({\bf q; kk'};t)
a^{\dag}_{{\bf k+q}}
a_{{\bf k}}b^{\dag}_{-{\bf k'-q}}b_{-{\bf k}'}, 
\end{eqnarray} 
is the effective {\em e--h} interaction; and 
\begin{eqnarray}
V_{ee}(t)= \frac{1}{2} \sum_{{\bf kk'q}}  
\upsilon_{ee}({\bf q; kk'};t) 
a^{\dag}_{{\bf k+q}}a^{\dag}_{{\bf k'-q}}a_{{\bf k'}}a_{{\bf k}},
\end{eqnarray} 
is the effective {\em e--e} interaction. The explicit
expressions for $\upsilon_{eh}({\bf q; kk'};t)$
and $\upsilon_{ee}({\bf q; kk'};t)$ are given
in Appendix C. As can be seen, $\tilde{H}(t)$ has the
same operator form as the bare Hamiltonian $H$. However,
both the effective band dispersions and the effective
interaction potentials are now {\em dependent on time}.
Note here that the above  pump--induced 
renormalizations  only last for the pulse duration 
$t_{p}$.
As discussed above, they are therefore perturbative 
in the screened interactions for $
t_p E_M\lesssim 1$ or for $\Omega > E_{M}$.

Let us first discuss the effect of the pump--induced
self--energy corrections to the conduction and
valence band energies, given by the last terms in Eqs.
(\ref{ed}) and (\ref{hd}). The dispersion of the
effective band is shown in Fig. 1. As can be seen,
the pump pulse leads to a bandgap increase as well as
a change in the momentum dependence (band dispersion) 
that last {\em as long as the pump pulse}. The
magnitude of the bandgap increase  is of the order of 
$(\mu{\cal E}_{p})^{2}/\Omega $ (for off--resonant
excitation) and $(\mu{\cal E}_{p})^{2} t_p$ (for
resonant excitation) and leads to, e.g., 
the ac--Stark blueshift. 
As we shall see, for pulse duration shorter than the
dephasing time, it also leads to bleaching and gain
right below the onset of absorption, analogous to
the case of excitons or two--level systems.\cite{stef,binder}  
It should be emphasized that these are {\em coherent}
effects that should not be confused with the incoherent
bandgap redshift due to the {\em e--e} interactions
among  real photoexcited carriers.\cite{wang} Note also that the
above bandgap renormalization is induced by the
{\em transverse} EM--field of the laser, as compared to the usual 
bandgap renormalization due to a {\em longitudinal}
EM--field, i.e., Coulomb screening.
The change in the band
dispersion, whose relative magnitude is of
the order of $(\mu{\cal E}_{p}/\Omega)^{2}$
(for off--resonant excitation) or
$(\mu{\cal E}_{p}t_{p})^{2}$ (for resonant
excitation), can be viewed as an increase in
the effective density of states and, to the
first approximation, in the effective mass.
This is important in doped semiconductors 
because, 
as we shall see later, it leads to a 
time--dependent  enhancement 
of  the {\em e--h} interactions and
scattering processes with the FS electrons.

The effective Hamiltonian $\tilde{H}(t)$ also includes
pump--induced corrections in the effective interaction
potentials, determined by the pair--pair and pair--FS
interactions during the pump photoexcitation. By expanding 
Eqs.\ (\ref{veh}) and (\ref{vee}) for carrier energies 
close to the Fermi surface using Eqs.\ (\ref{eeh}) 
and (\ref{ehh}), one can show that these corrections
vanish at the Fermi surface; for the typical FS excitation
energies $\Delta \varepsilon \sim E_M$ that contribute
to the FES, their order of magnitude is 
$(\mu{\cal E}_{p}\,\Delta \varepsilon /\Omega^{2})^{2}$ 
(for off--resonant  excitation) or 
$ (\mu{\cal E}_{p}\,\Delta\varepsilon\, t_{p}^{2})^{2}$ 
(for resonant excitation). Thus the corrections to the
interaction potentials are suppressed for below--resonant
excitation by a factor of $(E_M /\Omega)^{2}$, or for
short pulses by a factor of $( E_M t_{p})^{2}$, 
as compared to the self--energy corrections. Such a
suppression is due to the Pauli blocking effect and the 
screening, which leads to the  vanishing of the pump--induced
corrections to the interaction
potentials at the Fermi surface. Similarly, the
pump--induced indirect optical transition matrix
elements $M_{{\bf pp'k}}(t)$ are suppressed by the
same factor as compared to the  direct transition
matrix element $M_{\bf{p}}(t)$ [first term
in  Eq.\ (\ref{tme})], while they  contribute
to the pump--probe polarization only in 
the second order in the screened interactions.
Therefore, in the doped case,
the screened Coulomb interaction 
leads to subdominant parameter 
renormalizations to the effective Hamiltonian 
$\tilde{H}(t)$ 
and transition operator $\tilde{U}^{\dag}(t)$
for sufficiently 
short
 pump pulses or for off--resonant excitation.
In the following Section, we study the {\em e--h}
dynamics caused by the pump--induced self--energies
and direct transition matrix elements which lead
to the strongest nonlinearities in the case of
below--resonant or short pulse excitation.

\section{ELECTRON--HOLE DYNAMICS}

In this Section, we derive the final formulae for
the nonlinear pump--probe polarization of the
interacting system by applying the CCE to the
effective Hamiltonian $\tilde{H}(t)$
in order to treat the dynamical FS response. The CCE
equation Eq.\ (\ref{phi}) contains the operator
$S(t)$ described by a hierarchy of coupled
equations for the amplitudes $s, s_{2}, \cdots$,
defined by Eq.\ (\ref{S}). In the coherent case of 
interest here, the {\em e--e} scattering effects 
are suppressed, and the nonlinear absorption spectrum 
is dominated by the {\em e--h} interactions. This allows
us to use the dephasing time approximation for treating the  
probe--induced {\em e--e} scattering 
processes \cite{hawr1,ohtaka,sham1,hawr2},
in which case the above 
hierarchy terminates after $s$. Importantly, the  
{\em e--h} correlations (dynamical FS response) are 
still treated 
non-perturbatively, since they are determined by $s$.\cite{schon,ilias4,ilias5}
Then all FS--pair creation processes can be eliminated
explicitely from the rhs of Eq.\ (\ref{phi}), leading
to the following nonlinear differential equation
for the one--FS--pair scattering amplitude,\cite{schon,ilias4,ilias5}
$s({\bf p},{\bf k},t)$:
\begin{eqnarray} \label{de} 
i\frac{\partial s({\bf p,k},t)}{\partial t}-
\left[\varepsilon_{{\bf p}}^{c}(t) - 
\varepsilon_{{\bf k}}^{c}(t)\right] 
s({\bf p,k},t) =V \left[ 1 + \sum_{p' > k_{F}}
s({\bf p',k},t) \right]  
\left[ 1 - \sum_{k' < k_{F}}s({\bf p,k'},t) \right].
\end{eqnarray}
The {\em e--h} scattering processes described by
the above equation are sketched in Fig. 2a. 
Here $V$ is the s--wave component \cite{rev1} 
of the screened  
interaction \cite{ander1}, 
 approximated for simplicity  by its value 
at the Fermi energy.\cite{andrei,ohtaka,sham1} 
This neglects plasmon effects, which are
however small within the  frequency range of the FES.\cite{hawr3,hawr1}
Although this approximation is standard for the linear
absorption case, its justification for the transient spectra
requires more attention. Indeed, the
characteristic time for screening buildup is of the order 
$\Omega_p^{-1}$, where  $\Omega_p$ is the typical
plasma frequency corresponding to the FS. 
\cite{el-sayed94,banyai98,mahan2,hawr3}
This time is however shorter than the
typical pump duration $\sim$ 100fs  and dephasing time (which is of the order
of ps near the Fermi energy),\cite{wang,kim} 
so the Coulomb interactions can be  considered 
screened also for the nonlinear absorption case.
In Eq.\ (\ref{de}), we neglected the hole recoil energy
contribution to the excitation energy \cite{ilias5}  since 
$\varepsilon^{v}_{k_{F}}(t)-\varepsilon^{v}_{0}(t)\lesssim E_M$
due to a  sufficiently heavy hole mass.\cite{andrei} 
Note that, by increasing the hole effective mass, 
the pump--induced hole self--energy, Eq.\ (\ref{hd}), 
reduces the hole recoil energy and thus the  corresponding
broadening.\cite{andrei} In real samples, the
relaxation of the momentum conservation condition
due to the disorder will also suppress the hole
recoil broadening effects. 

>From  Eq.\ (\ref{phi}) we then easily obtain the
following expression for $|\Phi(t) \rangle$:
\begin{equation} \label{photo}
|\Phi(t) \rangle = \sum_{p>k_{F}} \Phi_{{\bf p}}(t,t') 
 a^{\dag}_{{\bf p}}b^{\dag} | 0 \rangle,
\end{equation}
where  $b^{\dag}$ is the creation operator of the 
zero--momentum hole state and the {\em e--h} pair
wavefunction $ \Phi_{{\bf p}}(t;t')$ satisfies the
``Wannier--like'' equation of motion,
\begin{eqnarray}\label{phip} 
 i \frac{\partial \Phi_{{\bf p}}(t,t')}{\partial t} 
= \left[\Omega + \varepsilon_{{\bf p}}^{c}(t) +
\varepsilon_{-{\bf p}}^{v}(t) - \epsilon_A (t) 
- i \Gamma_{{\bf p}}\right] \Phi_{{\bf p}}(t,t')
- \tilde{V}({\bf p}, t)\sum_{p' > k_{F}}
\Phi_{{\bf p'}}(t,t'),
\end{eqnarray}
where
\begin{equation}\label{Veff}
\tilde{V}({\bf p}, t) =
V\left[ 1 - \sum_{k'<k_{F}} s({\bf p,k'},t) \right],
\end{equation}
is the  effective    {\em e--h} 
potential whose 
{\em time-- and momentum--dependence}
is determined by the
response of the FS electrons to their 
 interactions with  the probe--induced
{\em e--h} pair (vertex corrections)
 [sketched schematically
in Fig. 2(b), 
 responsible for the
unbinding of the HFA bound state], 
\begin{equation} 
\epsilon_A (t)=V \sum_{k'<k_{F}} \left[1+
\sum_{p'>k_{F}} s({\bf p',k'},t)\right]
\end{equation} 
is the self--energy due the to the sudden appearance
of the photoexcited hole, 
which leads to non--exponential polarization
decay [described by ${\rm Im}\,\epsilon_A(t)$] due
to the  Anderson orthogonality catastrophe \cite{ander}  
and a dynamical resonance redshift [described by
${\rm Re}\,\epsilon_A(t)$], and $\Gamma_{{\bf p}}$
describes all additional  dephasing processes (due
to {\em e--e} interactions, hole recoil, and phonons).
Eq.\ (\ref{phip}) should be solved with the initial
condition $\Phi_{{\bf p}}(t',t') = M_{{\bf p}}(t')$, 
where $M_{{\bf p}}(t')$ is defined by Eq.\ (\ref{tme}). 

It is worth stressing here the analogy between Eqs.
(\ref{photo}) and (\ref{phip}) and the corresponding
problem in undoped semiconductors. Indeed, Eq. (\ref{photo})
is the direct analog of an exciton state, whereas
again Eq. (\ref{phip}) is very similar to a Wannier
equation. However, Eqs. (\ref{de}) and (\ref{phip})
include the effects of the interactions between the
probe--photoexcited 
{\em e--h} pair and the FS--excitations, and  the 
wavefunction $\Phi_{{\bf p}}(t,t')$ describes
the propagation of the photoexcited pair
``dressed'' by the FS excitations. Such a ``dressing'' 
is due to the dynamical FS response, which leads to  
the dynamical screening of the effective {\em e--h}
interaction 
Eq.\ (\ref{Veff}). The time--dependence of the latter  is determined
by the FS scattering amplitude $s({\bf p,k},t)$
and is affected by the pump excitation as described by Eq.\ (\ref{de}).
One can easily verify that
by setting $s({\bf p,k},t)=0$ in Eq.\ (\ref{phip}), we
recover the results of the Hartree--Fock (ladder diagram,
static FS\cite{andrei}) approximation. If one neglects
the nonlinear (quadratic) term in  Eq.\ (\ref{de}),
one recovers the three--body (Fadeev)
equations.\cite{andrei,fadeev} 
Note that the coupled equations
for $\Phi_{{\bf p}}(t,t')$ and $s({\bf p,k},t)$,
obtained by neglecting the multipair excitations
in Eq.\ (\ref{S}), can  be extended to include the
hole recoil--induced corrections.\cite{ilias5} 

Using Eqs. (\ref{ccw}), (\ref{S}), and (\ref{photo}),
we now can express the pump--probe polarization Eq.\
(\ref{spec1}) in terms of the {\em e--h} wavefunction
$\Phi$ and the effective transition matrix element
$M_{{\bf p}}$. Assuming, for simplicity, a
delta--function probe pulse centered at time
delay $\tau$, 
${\cal E}_s(t)={\cal E}_s\delta(t-\tau)$, we obtain
a simple    final expression
($t>\tau$):
\begin{eqnarray}\label{p-p} 
P_{{\bf k}_{s}}(t)= - i\mu^2{\cal E}_s
e^{i{\bf k}_{s}\cdot {\bf r}- i\omega_{p} (t - \tau)} 
\sum_{p > k_{F} } M_{{\bf p}}(t)  \Phi_{{\bf p}}(t,\tau). 
\end{eqnarray}
Eq.\ (\ref{p-p}) expresses the pump--probe
polarization in terms of two physically distinct
contributions. First is the effective transition
matrix element $M_{{\bf p}}(t)$, which includes
the effects of pair--pair and pair--FS interactions
and Phase space filling effects due to the
pump--induced carriers present during the probe
photoexcitation. Second is the  wavefunction
$\Phi_{{\bf p}}(t)$ of the {\em e--h} pair
photoexcited by the probe, whose time dependence, 
determined by Eqs.\ (\ref{phip}) and (\ref{de}), describes the
formation of the absorption resonance.
 Despite the formal similarities,
there are two important differences between the
doped and the undoped cases. First, in the doped
case, the time evolution of the {\em e--h}
wavefunction $\Phi_{{\bf p}}(t)$ is strongly
affected by the interplay between the {\em e--h}
correlations and the pump--induced 
transient changes in
the bandgap and band dispersion relations.
As we shall see later, this can be viewed as 
an excitation--induced dephasing. Second,
unlike in the undoped case,\cite{per99} the
pump--induced corrections in the effective
matrix element $M_{{\bf p}}(t)$ are
perturbative in the screened interactions
if the pump detuning or the pump frequency
width exceed the Coulomb energy $E_M$. In
the next Section, we demonstrate the role
of each of these effects both analytically
(for CW excitation) and numerically (for
short pulses).

\section{ PUMP--PROBE DYNAMICS} 
\subsection{Monochromatic Excitation} 

In this subsection, we analyze the FES pump--probe
spectrum in the case of monochromatic excitation. 
For monochromatic pump, the theory developed in
the previous Sections applies for  pump detunings
larger than the characteristic Coulomb energy, 
$\Omega\gtrsim E_M$. Close to the Fermi edge, the
linear absorption spectrum of the FES can be
approximately described with the analytic expression
\cite{andrei,xray} 
\begin{equation}  \label{fes0} 
\alpha(\omega) \propto  {\cal N}
\left( \frac{E_{F}}{\omega}\right)^{\beta}, 
\end{equation}
where ${\cal N}$ is  the density of states, $\omega$ 
is the optical frequency measured  from the Fermi
edge, and $\beta=2 \delta/\pi- \left(\delta/\pi \right)^{2}$
is the FES exponent, where $\delta \sim  \tan^{-1}(\pi g)$ 
is the s--wave phaseshift of the screened {\em e--h}
potential evaluated at $E_F$, and $g={\cal N}V$ is
the dimensionless parameter characterizing the scattering
strength.  The monochromatic
pump excitation leads to a resonance blueshift,
originating from the shift in the effective band
energies (see Fig. 1), and to a bleaching mainly
due to the Pauli blocking which reduces the effective
transition matrix element (analogous to the dressed
atom picture \cite{cohen}). More importantly, however,
the pump--induced change in the band dispersion
increases the density of states
 ${\cal N}$  close to the Fermi
surface and thus also increases both the {\em e--h}
{\em scattering} strength $g$ and the phaseshift $\delta$.
This, in turn, leads to an increase in the FES
{\em exponent} $\beta$ that determines the resonance
{\em width} and lineshape. In contrast, in the case
of a bound excitonic state of dimensionality $D$
and Bohr radius $a_{B}$, the resonance width remains
unchanged, while the oscillator strength,
$\propto a_{B}^{-D}$, increases by a factor
$ \sim (1 + D\,\Delta m)$, where $\Delta m$ is the
pump--induced change in the effective
mass \cite{ilias2}. Such an optically--induced enhancement
of the exciton strength competes with the bleaching
due to the Pauli blocking and exciton--exciton
interactions. This
results in an almost rigid exciton blueshift, consistent  
with experiment \cite{knox1,chemla89} and previous
theoretical results \cite{ell}. 

However, in the case of a FES resonance, the
pump--induced change in the exponent $\delta$ leads
to a stronger oscillator strength enhancement than
for a bound exciton state. Obviously, such an
enhancement cannot be described perturbatively,
i.e., with an expansion in terms of  the optical
field, since the corresponding corrections to the
absorption spectrum diverge logarithmically at the
Fermi edge. 
As can be seen from Eq.\ (\ref{fes0}), 
the effect of the pump on the FES can
be thought of as  an excitation--induced dephasing
that affects the {\em frequency dependence} of
the resonance; again, this is in contrast to the
case of the exciton. In the time domain,
this also implies a memory structure 
related to the response--time of the FS-excitations.
Therefore, the qualitative differences between
the nonlinear optical response of the FES and 
the exciton
originate from the fact that an exciton is a discrete
bound state, while the FES is a {\em continuum}
many--body resonance. The FS responds {\em
unadiabatically} to the pump--induced change in
the density of states via an {\em increase} in
the {\em e--h}
scattering of {\em low--energy} pair excitations.
Such  scattering processes, which determine the
response of the Fermi sea to the hole potential
in the course of the optical excitation,
are responsible for the unbinding 
and broadeningof the HFA bound state.\cite{mahan2} 
Therefore, the pump field 
changes the broadening and dephasing effects
even for  below--resonant photoexcitation. 
On the other hand, due
to the finite Coulomb binding energy of the exciton, 
the pump optical field can polarize such a bound
state and change its Bohr radius without ionizing
it. In the next subsection, we study the above 
effects in the case of  short pulse excitation.

\subsection{Short--pulse Excitation} 

We now present our results for the nonlinear
absorption of the FES in the case of short pulse
excitation. The results were  obtained by solving
numerically the differential equations
(\ref{phip}) and (\ref{de}), using the Runge--Kutta
method, for Gaussian pulses with duration
$t_p = 2.0 E_F ^{-1}$. We only considered
negative time delays and  focussed on
below--resonant pump excitation in order
to suppress the incoherent effects due to the
{\em e--e} scattering of real pump--induced
carrier populations with the FS. Under such
excitation conditions, the {\em coherent}
effects in which we are interested dominate, 
and the Coulomb--induced corrections 
to the effective  parameters, discussed
in Section IV, are perturbative.  Our
goal is to study the role of the dynamical FS
response ({\em e--h} correlations) on the
pump--probe dynamics. For this reason we compare
the results of our theory to those of the HFA,
obtained by setting $s({\bf p},{\bf k},t)=0$
in Eq.\ (\ref{phip}). As mentioned above, in
the latter case, 
the (spurious) HFA bound state does
not interact with the FS pair excitations, even
though it can merge with the continuum when one
introduces a very short dephasing time. 

In Fig. 3, we compare the linear absorption lineshape 
(in the absence of pump, ${\cal E}_p(t) = 0$) of the FES
to the HFA (without the dynamical FS response). 
We use the parameter values $g=0.4$ and 
$\Gamma=0.1 E_{F}$, which were 
previously used to fit the experimental spectra 
in  modulation doped quantum wells.\cite{ohtaka,xray} 
For better visibility, we shifted
the
curves in order to compare their lineshapes.
The linear absorption FES lineshape is consistent with
that obtained in Refs. \onlinecite{xray} and
\onlinecite{ilias4}. On the other hand, the HFA
spectrum is characterized by the coexistence of
the bound state and a continuum contribution
due to the fact that, in 2D, a bound state exists
even for an arbitrary weak attractive potential.
We note that if one limits oneself to linear absorption,
it is possible to artificially shorten the dephasing
time $T_2 = \Gamma^{-1}$, mainly determined by the hole
recoil effects, by taking $\Gamma \simeq E_M$. Then the
spurious discrete state and the continuum merge, and
the discrepancy between the two linear absorption
lineshapes decreases. This trick has been
used for phenomenological fits of linear absorption 
experimental data.\cite{fes} Below we show, however, 
that in the nonlinear absorption case, the differences in the
transient spectra are significant so that the processes 
beyond HFA can be observed experimentally.

Let us turn to the time evolution of the pump--probe
spectra. In Fig. 4 we show the nonlinear absorption
spectra calculated by including the dynamical FS response 
(Fig. 4(a)) and within the HFA (Fig. 4(b))
at a short time delay $\tau = - t_p/2$.
The main features of the spectrum 
are a pump--induced resonance bleaching, blueshift,
and gain right below the onset of absorption. For
off--resonant pump, these transient effects vanish
for positive time delays after the pump is gone, and
persist for negative time delays shorter than the
dephasing time $T_2 = \Gamma^{-1}$. Similar features
were also obtained for different values of the pump
amplitude, duration, and  detuning. They are mainly due to
the broadening induced by the transient renormalization
of the energy band dispersion [Eqs.\ (\ref{ed}) and
(\ref{hd})] when its duration $\sim t_{p}$ is
shorter than the dephasing time (analogous to
excitons and  two--level systems\cite{stef,binder}).

We now turn to the role of the {\em e--h}
correlations. In Fig. 5 we compare the differential
transmission spectrum calculated by including the dynamical
FES response or within the HFA for long and short 
negative time delays. Note that, in  pump/probe
spectroscopy,
the experimentally measured differential transmission is 
\begin{equation} \label{diff-T}
DST (\omega, \tau) =
\frac{\Delta T_s(\omega, \tau)} {T_0 (\omega)}
\, = \, \frac{ T_{s}({\cal E}_p)
-T_{s}({\cal E}_p=0)}
{T_{s}({\cal E}_p=0)} \, ,
\end{equation}
where $T_{s}({\cal E}_p)$ is the transmission coefficient in the
probe direction in the presence of the pump field ${\cal E}_p$.
In the weak signal regime, it reproduces the changes
in the probe absorption coefficient
$\alpha (\omega , \tau)$:
$DST(\omega,\tau)\propto-\Delta \alpha(\omega ,\tau)$.
Fig. 5(a) shows 
the results obtained for a  long time delay,
$\tau = -1.5T_2= - 15.0E_F^{-1}$,
in which case frequency
domain oscillations are observed. 
These oscillations are
similar to those seen in undoped semiconductors
and two--level systems; however, their amplitude
in the FES case is reduced.  On the other hand, 
as shown in Fig. 5(b),
for time delays comparable to the pulse duration,
$\tau = - 0.1 T_2 = - t_p/2 = - 1.0 E_F^{-1}$,
the main features are a blueshift and bleaching. 
In this case the {\em e--h} correlations lead to
a substantially larger width and asymmetric lineshape
of the differential transmission spectrum. This comes
from the different
response of the FES 
to the pump--induced dispersion renormalizations 
when the {\em e--h} correlations are accounted for. 
 This is
more clearly seen in Fig. 6(a), where we plot the
magnitude of the resonance decrease, evaluated at
the peak frequency, as a function of $\tau$. We use, of course,
the same values of the parameters in the two calculations
and yet we find that the bleaching of the FES peak
is substantially stronger when
the dynamical FS response is included 
than in the HFA case. Note  that for 
$|\tau| \sim \Gamma^{-1}$ the
FES resonance is actually enhanced by the pump, 
as can be seen more clearly in Fig. 7. 
The time dependence of the resonance bleaching is
strikingly different in the two cases. In the HFA
case,  $ |DST(\omega,\tau)|$ evaluated at the 
instantaneous  peak frequency decays over
a time scale $|\tau| \sim \Gamma^{-1}$, i.e. during
the dephasing time. This is similar to results
obtained for a two--level system with the same
effective parameters. On the other hand, the decay
of $|DST(\omega,\tau)|$ at the peak frequency
is much faster when we take into account the {\em e--h}
correlations. Note that the above results were obtained for
off--resonant  excitation. Under
resonant conditions, we find that a spectral hole is produced. 
In Fig. 8 we compare the resonance blueshifts, 
evaluated at the peak frequency, as a function of
$\tau$. Again, a larger blueshift is predicted when 
the dynamical FS response 
is included. This suggests that
in the experiment of Ref.\ \onlinecite{igal}, where
similar blueshifts were observed in two quantum well
samples (one modulation doped with a FES and one
undoped sample with a 2D-exciton) the effective
parameters were larger in the latter case, due to
the absence of screening and exciton--exciton
interaction effects. 

In order to gain qualitative understanding of the role of the 
{\em e--h} correlations, let us
for a moment neglect the momentum dependence of the
pump--renormalization of the band dispersion and
the phase space filling effects and  consider the bleaching 
caused by a  rigid semiconductor band shift $\Delta E_{g}(t)$,
obtained from the
pump--induced self--energies, Eqs. (\ref{ed})
and (\ref{hd}), evaluated  at the bottom of the
band (note that $\Delta E_{g}(t)$ lasts for the duration of pulse). 
Within this model, the pump
excitation has no effect on the {\em e--h}
scattering amplitude $s({\bf p},{\bf k},t)$
[see Eq.\ (\ref{de})]. It is thus convenient to factorize 
the effects of the rigid band shift on the {\em e--h} 
wavefunction $\Phi_{{\bf p}}(t,t')$:
\begin{equation}
\label{factor}
\Phi_{{\bf p}}(t,t')=
e^{- i \int_{t'}^{t} \Delta E_{g}(t'') dt''}
 \, \tilde{\Phi}_{\bf p}(t,t') \, .
\end{equation}
This relation is general and defines 
$\tilde{\Phi}_{{\bf p}}(t,t')$, which does {\em not} 
depend on $\Delta E_{g}(t)$. In the special 
case of a rigid shift, 
$\tilde{\Phi}_{{\bf p}}(t,t')$ coincides
with $\Phi_{{\bf p}}^0(t-t')$ describing  the
propagation of the probe--photoexcited {\em e--h} pair 
in the {\em absence} of the pump pulse.
By substituting into Eq.\ (\ref{factor}) the long--time
asymptotic expression\cite{mahan2} 
$\tilde{\Phi}_{{\bf p}}(t,t') 
=\Phi_{{\bf p}}^0(t-t')\propto[i(t-t') E_{F}]^{\beta-1}$ that
gives the linear absorption spectrum of the FES at 
$\omega \simeq E_F$, and substituting the resulting 
$\Phi_{{\bf p}}(t,t')$ into Eq.\ (\ref{p-p}), 
we obtain a simple  analytic expression
for the effect of a pump--induced
 rigid band shift on the nonlinear 
absorption spectrum: 
\begin{equation}\label{nla} 
\alpha(\omega) \propto {\rm Re}\,
\int_{\tau}^{\infty} dt e^{i(\omega +i\Gamma)
(t-\tau)- i \int_{\tau}^{t}
\Delta E_{g}(t') dt' } 
[i(t-\tau) E_{F}]^{\beta-1}.
\end{equation}
For $\Delta E_{g}(t)=0$ one, of course, recovers
the linear FES absorption in the vicinity of the 
Fermi edge.\cite{mahan2} 
For $\beta=0$, 
 Eq.\ (\ref{nla}) gives the absorption of 
the non--interacting continuum. 

The physics of the FES can be seen from Eq.\ (\ref{nla}). 
For $\beta=1$, this gives a discrete 
Lorentzian peak corresponding to the HFA bound state.  
However, during the  optical transition, the {\em e--h} 
pair interacts with the FS electrons, leading to the 
and the readjustment of the FS density profile 
via the scattering of FS pairs.
This results in the  broadening of the discrete HFA bound state,  
which is governed  by the time evolution of the FS. 
Such time evolution is unadiabatic due to the low--energy FS pairs,
which leads to the characteristic power--law time dependence of the
broadening factor in Eq.\ (\ref{nla}). The interaction with the
FS-pairs determines the exponent, $0 \leq \beta \leq 1$,
of the latter, which leads to a non--Lorentzian
lineshape in the frequency domain and a
non--exponential decay in the time domain. 
A detailed discussion of the above physics and the analogy 
to phonon sidebands and collision broadening may be found in 
Ref. \onlinecite{mahan2}. 
Here it is essential to use the CCE in order to calculate the spectrum at
{\em all} frequencies 
(and not just asymptotically close to $E_{F}$ as 
with Eq.\ (\ref{nla})) 
and, most importantly, 
to describe the non--equilibrium FS and {\em e--h} pair response to the
time--dependent increase in the 
effective mass/density of states, not included in 
Eq.\ (\ref{nla}).

The resonance bleaching obtained from 
Eq.\ (\ref{nla}) as a function of $\tau$, is shown in Fig. 6(b) 
for $\beta=0.6$, corresponding to the value of the parameters used 
in Fig. 6(a) ($g=0.4$), together with the HFA
result ($\beta =1$). Comparing Figs. 6(a) and (b),
one can see that the rigid band shift approximation
qualitatively accounts for the dynamics, but that
there are strong  discrepancies (see vertical scales), 
whose origin is discussed below. Both the magnitude and the
time--dependence of the bleaching depends critically
on the value of $\beta$, which characterizes the 
interaction of the photoexcited {\em e--h} pair 
with the FS excitations. Because of such
coupling, many polarization components are
excited in the case of the continuum FES resonance, 
and it is their interference that governs the 
dynamics of the pump/probe signal. 
Such interference 
is  also responsible for the resonance enhancement and
differential transmission oscillations at $\tau <0$
shown in Figs. 5 and 7. As $\beta$ increases the
interference effects are suppressed because the  
energy  width of the continuum states contributing
to the FES narrows. In fact, this energy width is 
directly related to that of the linear absorption resonance.
This is clearly seen in Fig. 9 where we show
the effect of increasing $g$ on the dynamics of the
bleaching. It becomes more bound--state--like as, 
with increasing $g$, the FES resonance becomes narrower. 
On the other hand, in the HFA case,
the decay rate is $\sim T_2 = \Gamma^{-1}$,
i. e. it is independent on $g$, when $E_M$ becomes
smaller than $\Gamma$, while for $E_M \simeq \Gamma$, 
the contribution of the continuum states produce
a faster decay.

Although the transient rigid band shift approximation,
Eq. (\ref{nla}), explains some of the features of
the dynamics of the bleaching, it strongly overestimates
its magnitude. This is because Eq.\ (\ref{nla})
neglects the response of the many--body system
to the pump--induced renormalization of the 
band's dispersion, 
neglected in Eq.\ (\ref{nla}).
 Such a  transient change in the 
dispersion, which can be viewed as an increase in the
density of states/effective mass
 for the duration of the pump,
is important because it results in an 
enhancement of  the {\em e--h} scattering. 
For example, 
in the case of monochromatic excitation, 
this leads to the change in the exponent $\beta$ 
of the broadening prefactor in the integrand of 
Eq.\ (\ref{nla}), discussed in the previous  
subsection. For the short--pulse case, 
it is not possible to describe analytically
the effect of the pump on the {\em e--h} 
scattering processes, 
due to the non--equilibrium unadiabatic FS response.
The latter can be described with the numerical 
solution of Eqs.\ (\ref{de}) and (\ref{phip}), 
which is
presented in Figs. 10 and 11 and discussed below.

In order to show  the role 
of the pump--induced renormalization of the band 
dispersion in the presense of the 
dynamical FS response, 
we plot in Fig. 10 the function 
\begin{equation} \label{F}
F(\omega,\tau) 
= {\rm Im} \sum_{p > k_{F}}
\tilde{\Phi}_{{\bf p}}(\omega ,\tau)
\end{equation}
where $\tilde{\Phi}_{{\bf p}}(\omega ,\tau)$ is the
Fourier transform of $\tilde{\Phi}_{{\bf p}}(t,\tau)$ defined by
Eq.\ (\ref{factor}). Note that, in the presence of the
band dispersion renormalization, the
wave--function $\tilde{\Phi}_{{\bf p}}$ (which is  independent of 
$\Delta E_g(t)$) no longer coincides with $\Phi_{\bf p}^0$
as in Eq.\ (\ref{nla}). As can be seen in Fig. 10(a), 
when the {\em e--h} correlations are taken into account,
 the  pump--induced redistribution
of oscillator strength between the states of the
continuum that contribute to the resonance manifests
itself as a dynamical redshift. This shift opposes
the rigid band blueshift $\Delta E_{g}(t)$ 
(when the latter is included). 
At the same time, the resonance strength is enhanced
significantly.  
The latter effect originates from the interplay between the 
transient increase in the effective mass/density of states 
of the photoexcited {\em e--h} pair
and the ``dressing'' of this pair  with the FS excitations
[described by the effective
potential $\tilde{V}({\bf p},t)$ in Eq.\ (\ref{phip})]. 
In contrast, such an oscillator strength enhancement is 
suppressed in the HFA (which neglects the {\em e--h} 
correlations), as seen in Fig. 10(b), 
in which case  the main feature
is the redshift of the resonance
due to the  pump--induced increase of the binding energy
$E_M$ coming from the transient increase in the effective
mass [see Fig. 10(b)]. 

In Fig. 11 we show the effect of the renormalization
of the band dispersion on the nonlinear absorption
spectrum. The optically--induced
increase in the {\em e--h} interactions
enhances significantly the strength of the FES
and compensates part of the bleaching induced by the
rigid band shift. A smaller enhancement is also seen
in the HFA, where the pump--induced increase in 
the binding energy $E_M$ competes with the effects
of the bandgap renormalization.

\section{Conclusions }

In summary, we developed a theory 
for the ultrafast nonlinear optical response of the FES.
We focussed on coherent effects, 
which dominate the  pump--probe spectra
during negative time delays 
and off--resonant excitation conditions. 
We  demonstrated  that the dynamical FS response 
leads to qualitatively different coherent  dynamics
of the FES  as compared to the Hartree--Fock
approximation.
In particular, in the former case, the time evolution of the
resonance bleaching is governed by the dephasing time,
while in the former case  polarization interference
effects dominate. This results in faster FES dynamics
(which depends on the strength of the screened 
{\em e--h} potential)
as well as an apparent resonance enhancement during
negative time delays. Such dynamical features  should be observable 
in ultrafast pump/probe experiments.
Using a simple model, we showed that the different
dynamics of the FES and Hartree-Fock treatment can
be attributed to the non--Lorentzian broadening of
the HFA bound state due to its interactions 
with the gapless FS excitations, a process which is,
of course, beyond the dephasing time approximation.
In addition, we showed that the pump excitation 
directly affects the strength of the {\em e--h} 
scattering processes, which changes the frequency
dependence of the resonance. The latter can be
thought of as an excitation--induced dephasing
effect that leads to a transient enhancement of
the FES. Our results indicate that ultrafast 
spectroscopy provides a powerful tool to study
the role of  correlations in the nonlinear response
of a Fermi liquid to the optical excitation during
time scales shorter than the dephasing times. 
The dynamical features discussed above can also be used 
as an experimental signal  to probe the
crossover from FES to exciton bound state 
(exciton Mott transition) as a function of the  FS density.

\acknowledgements

This work was supported by the NSF Grant No. ECS--9703453,
by  a grant from the  HITACHI Advanced Research Laboratory, and in
part  by the Office of Naval
Research grant No. N00014-96-1-1042.
The work of D.S.C. was supported by the Director, Office
of Energy Research, Office of Basic Energy Sciences, Division
of Material Sciences of the U.S. Department of Energy,
under Contract No. DE-AC03-76SF00098.


\appendix
%
%
%
\section{}

In this appendix we briefly outline our formalism.\cite{per99} 
The pump--probe signal is determined by the polarization,
\begin{equation}\label{polar}
P(t) = \mu e^{-i\omega_{p}t}\langle \Psi(t) | U |
\Psi(t) \rangle, 
\end{equation}
where the state $|\Psi(t)\rangle$ satisfies the
time-dependent Schr\"{o}dinger equation,
\begin{equation}\label{schro-1}
\left[i \frac{\partial}{\partial t} - H_{tot}(t)\right]
|\Psi(t) \rangle = \left[i \frac{\partial}{\partial t} -
H -H_{p}(t)-H_{s}(t)\right] |\Psi(t) \rangle =0 \, , 
\end{equation}
with the Hamiltonians $H$ and $H_{p,s}(t)$,
given by Eqs. (\ref{H0})--(\ref{Hp}).
The Hilbert
space of the bare semiconductor, i.e. in the absence
of optical fields,
consists of disconnected subspaces
$\zeta \{ \nu_{eh} \}$ which are labeled by the
number of (interband) {\em e--h} pairs, $\nu_{eh}$.
The corresponding bare Hamiltonian, $H$, conserves
the number of {\em e--h} pairs in each band separately
and in the $\zeta \{ \nu_{eh} \}$-basis has a
block--diagonal form. The Hamiltonians $H_{p,s}(t)$
couple the different subspaces $\zeta \{ \nu_{eh} \}$
by causing interband transitions.

In the description of pump/probe experiments, we
are interested only in the polarization component
propagating along the probe direction ${\bf k}_{s}$.
For a weak probe, the nonlinear signal
then arises from the linear response of the
pump/bare--semiconductor coupled system, described
by the time--dependent Hamiltonian $H + H_{p}(t)$,
to the probe--induced  perturbation $H_s(t)$.
Within $\chi^{(3)}$, the above is  true 
even for comparable pump and probe  amplitudes.
However, since, in contrast to $H$, the Hamiltonian $H + H_{p}(t)$ 
does not conserve the number of carriers in each band
separately,  the calculation of the linear response
function is not practical. Therefore, we seek 
to replace $H + H_{p}(t)$ by
an effective Hamiltonian $\tilde{H}(t)$
that {\em does} conserve the number of {\em e--h} pairs in
each of its Hilbert subspaces (i.e., is
``block--diagonal''). 
As derived in Ref. \onlinecite{per99}, 
this can be accomplished
in any given order
in the pump field ${\cal E}_p(t)$
and for any pulse duration 
by using  
a time--dependent Schrieffer--Wolff /Van Vleck
canonical transformation \cite{s-w,cohen}.
Here it is sufficient to 
``block--diagonalize''
the Hamiltonian $H+ H_{p}(t)$ up to the second order
in  ${\cal E}_p(t)$. 
The  transformation that 
achieves this 
has the form $e^{-\hat{T}_{2}}
e^{-\hat{T}_{1}}[H+H_{p}(t)]e^{\hat{T}_{1}}e^{\hat{T}_{2}}$,
where the anti--Hermitian operators $\hat{T}_{1}(t)$
and $\hat{T}_{2}(t)$ create/annihilate one and two
{\em e--h} pairs, respectively.

We proceed with the first step and eliminate the
single-pair pump-induced transitions in the
time--dependent Schr\"{o}dinger equation of the
pump/bare--semiconductor system,
\begin{equation} 
\left[i \frac{ \partial}{ \partial t} -H-H_{p}(t)\right]
|\Psi(t) \rangle=0 \, .
\label{schro-2}
\end{equation}
This is achieved by substituting
$|\Psi(t) \rangle = e^{\hat{T}_1(t)}|\chi(t) \rangle$
and acting with the operator $e^{-\hat{T}_1(t)}$
on the lhs of Eq.\ (\ref{schro-2}),
\begin{equation} 
e^{-\hat{T}_1(t) }
\left[ i \frac{\partial}{\partial t}
- H \right] e^{\hat{T}_1(t) } | \chi(t) \rangle 
=  e^{-\hat{T}_1(t) } \left[ H_{p}(t) \right] 
 e^{\hat{T}_1(t) } | \chi(t) \rangle.
\label{schroed}
\end{equation}
The anti--Hermitian operator $\hat{T}_{1}(t)$ has
a decomposition 
\begin{equation} 
\hat{T}_{1}(t) = \hat{\cal P}(t) 
e^{-i {\bf k}_{p} \cdot {\bf r}} -
\hat{\cal P}^{\dag}(t) e^{i {\bf k}_{p} \cdot {\bf r}},
\label{unit} 
\end{equation}
where $\hat{\cal P}^{\dag}(t)$ and $\hat{\cal P}(t)$
create and annihilate single {\em e--h} pairs,
respectively.
The effective Hamiltonian
can be found from the condition that the terms
describing single--pair interband transitions
cancel each other
in  Eq.\ (\ref{schroed}). 
In Ref. \onlinecite{per99}, 
it was shown that the multiple commutators 
of $\hat{\cal P}(t)$ with its time derivatives can be eliminated 
from  Eq.\ (\ref{schroed}) 
to all orders. 
By expanding Eq.\ (\ref{schroed})
and neglecting third or higher 
order terms in $\hat{\cal P}$, 
we obtain the following equation: 
\begin{equation}
i\frac{\partial \hat{\cal P}^{\dag}(t)}{\partial t} = 
\left[ H,\hat{\cal P}^{\dag}(t)\right] 
- \mu {\cal E}_p(t) U^{\dag},
\label{sigma1}
\end{equation}
with initial condition
$\hat{\cal P}^{\dag}(-\infty)=0$. The formal
solution is
\begin{equation} 
\hat{\cal P}^{\dag}(t)
= i\mu  \int_{-\infty}^{t} dt'{\cal E}_{p}(t')
e^{ -i H (t-t')} U^{\dag} e^{iH(t-t')}.
\label{sigop}
\end{equation}
Note that, since the Hamiltonian $H$ conserves the
number of {\em e--h} pairs and the optical transition
operator $U^{\dag}$ creates a single {\em e--h} pair, 
$\hat{\cal P}^{\dag}(t)$ also creates a single
{\em e--h} pair. Furthermore, since both $H$ and
$U^{\dag}$ conserve momentum, so does 
$\hat{\cal P}^{\dag}(t)$. 
Eq.\ (\ref{schroed}) then takes the form
\begin{eqnarray}
\left[i   \frac{\partial}{\partial t} 
-\tilde{H}(t)\right]|\chi(t) \rangle  = 
- \frac{\mu}{2}
\left[{\cal E}_p(t)
e^{2 i {\bf k}_{p} {\bf r}} 
\left[U^{\dag}, \hat{\cal P}^{\dag}(t) \right]
+ {\rm H.c.} \right]
| \chi(t) \rangle,
\label{first}
\end{eqnarray}
where 
\begin{equation} 
\tilde{H}(t) = H - \frac{\mu}{2}\,\left({\cal E}_p(t)
\left[\hat{\cal P}(t),U^{\dag} \right]
+ {\rm H.c.} \right)
 \label{eff1} 
\end{equation} 
is the sought time--dependent effective
Hamiltonian that  conserves the number of {\em e--h}
pairs and $\hat{\cal P}^{\dag}(t)$ is given by 
Eq.\ (\ref{sigma1}). Note that, since $\hat{\cal P}^{\dag}(t)$
is linear in the 
pump field ${\cal E}_p$, the pump--induced
term in $\tilde{H}(t)$ [second term in Eq.\ (\ref{eff1})] is quadratic
($\propto {\cal E}_p{\cal E}_p^{\ast}$). The rhs
of Eq.\ (\ref{first}) describes the pump--induced
two--pair transitions. These can be eliminated as well
by performing a second canonical transformation, 
$| \chi(t)\rangle = e^{\hat{T}_{2}(t)} |\Phi(t) \rangle$.
Following the same procedure, 
we use the anti--Hermiticity of $\hat{T}_{2}(t)$ to decompose it as 
\begin{equation} 
\hat{T}_{2}(t) = \hat{\cal P}_{2}(t) 
e^{-2 i {\bf k}_{p} \cdot {\bf r}} -
{\cal P}^{\dag}_{2}(t) e^{ 2i {\bf k}_{p} \cdot {\bf r}},
\label{unit2} 
\end{equation}
where ${\cal P}^{\dag}_{2}(t)$ and $\hat{\cal P}_{2}(t)$
create and annihilate two {\em e--h} pairs, respectively.
Substituting the above expression for $|\chi(t)\rangle$
into Eq.\ (\ref{first}) and requiring that all two--pair
transitions  cancel out, we obtain the following
equation for ${\cal P}^{\dag}_{2}(t)$,
\begin{equation}
i\frac{\partial {\cal P}^{\dag}_{2}(t)}{\partial t} = 
\left[ \tilde{H}(t),{\cal P}^{\dag}_{2}(t) \right] - \frac{\mu}{2}\,
{\cal E}_p(t) [\hat{\cal P}^{\dag}(t),U^{\dag}].
\label{sig2}
\end{equation}
Note that $\hat{\cal P}_{2}^{\dag}(t)$ only affects
the pump/probe polarization via  higher order
(${\cal E}_{p}^{4}$) corrections, which are neglected here.
However, it does determine the four--wave--mixing
(FWM) polarization (see below).

To obtain the condition of validity of this approach,
it is useful to write down a formal solution
(\ref{sigop}) of Eq.\ (\ref{sigma1}) in the basis
of the N--hole many--body eigenstates,
$|\alpha N \rangle$, with energies $E_{\alpha N}$, 
of the Hamiltonian $H$. Here the index $\alpha$
labels all the other quantum numbers, so that
$N$=0 corresponds to the semiconductor ground state
$| 0 \rangle$, $|\alpha 1\rangle$ denotes the one--pair states
(exciton eigenstates in the undoped case, with $\alpha$ labeling
both bound and scattering states), $|\alpha 2\rangle$ denotes the
two--pair (biexciton in the undoped case) eigenstates, etc. 
In this basis, the solution of Eq.\ (\ref{sigma1}) can be written as
\begin{equation}
\frac{  \langle \beta N+1
|\hat{\cal P}^{\dag}(t) |\alpha N \rangle}{
\langle \beta N+1| U^{\dag} |\alpha N \rangle }
=i\mu\int_{-\infty}^{t}dt' {\cal E}_{p}(t')
e^{-i(t-t') \left( \Omega
+ \Delta E_{\alpha \beta}^{N} \right)}
e^{-\Gamma(t-t')},  \label{sigm1}
\end{equation}
where we separated out the detuning $\Omega$ and denoted
$\Delta E_{\alpha \beta}^{N}=E_{\beta N+1}-E_{\alpha N}$.
It can be seen that for resonant excitation (small
$\Omega$) the rhs of Eq.\ (\ref{sigm1}) is of the order
of $\mu {\cal E}_{p} t_{p}$. Thus, for short pulses,
this parameter justifies the expansion in terms of
the optical fields. For off-resonant excitation, this
expansion is valid for any pulse duration provided
that $\mu {\cal E}_p/\Omega\lesssim 1$. Similar
conditions can be obtained for the two--pair transition
described by $\hat{\cal P}_{2}$.

The nonlinear polarization Eq.\ (\ref{polar}) can now
be expressed in terms of the linear response
to the probe field:
\begin{equation}\label{polar-p}
P(t) = \mu e^{-i\omega_{p}t}\langle \Psi(t) | U |\Psi(t)\rangle
= \mu e^{-i\omega_{p}t}\langle \Phi(t)|U_{T}(t) |\Phi(t)\rangle,
\end{equation}
where, in the first order in ${\cal E}_{s}(t)$, the
state $| \Phi(t)\rangle$ is given by
\begin{equation} 
|\Phi(t) \rangle = |\Phi_0(t)\rangle 
- i\mu\int_{-\infty}^{t}dt'{\cal K}(t,t') 
\left[  {\cal E}_{s}(t')e^{i  {\bf k}_{s}
{\bf r}+i\omega_{p}\tau} U^{\dag}_{T}(t') 
+ {\rm H.c.} \right] |\Phi_0(t')\rangle.
\label{phipert}
\end{equation}
Here ${\cal K}(t,t')$ is the time-evolution
operator satisfying
\begin{equation}
i \frac{\partial}{\partial t}{\cal K}(t,t')=
\tilde{H}(t){\cal K}(t,t'), \label{evol} 
\end{equation}
and
$U^{\dag}_{T}(t) = e^{-\hat{T}_{2}(t) }
e^{-\hat{T}_{1}(t) } U^{\dag}
e^{\hat{T}_{1}(t)}e^{\hat{T}_{2}(t) }$
is the (transformed) optical transition
operator. In Eq.\ (\ref{phipert}),
$|\Phi_0(t)\rangle = {\cal K}(t,-\infty)
|0 \rangle$ is the time--evolved ground state 
$|0 \rangle$; since $\tilde{H}(t)$ conserves the
number of {\em e--h} pairs,  
$|\Phi_0(t)\rangle$ contains no interband {\em e--h} pairs (in
undoped semiconductors, it coincides with the ground state, 
$|\Phi_0(t)\rangle= |0\rangle$). 
>From  Eqs.\ (\ref{phipert}) and (\ref{polar-p}),
the polarization $P(t)$ takes the form 
\begin{eqnarray} \label{pol}
P(t)&& =-i\mu^2 e^{-i\omega_{p}t}\int_{-\infty}^{t}dt'
\nonumber\\ &&\times
\Biggl[\langle\Phi_0(t)| U_{T}(t){\cal K}(t,t')
[{\cal E}_{s}(t') e^{i{\bf k}_{s}\cdot{\bf r}
+i\omega_{p}\tau} U^{\dag}_{T}(t')
+{\cal E}_{s}^{\ast}(t')e^{-i{\bf k}_{s}\cdot{\bf r}
-i\omega_{p}\tau} U_{T}(t')] |\Phi_0(t')\rangle
\nonumber\\&&
-\langle\Phi_0(t')|
\left[{\cal E}_{s}(t')e^{i{\bf k}_{s}\cdot{\bf r}
+i\omega_{p}\tau} U^{\dag}_{T}(t') 
+{\cal E}_{s}^{\ast}(t')e^{-i{\bf k}_{s}\cdot{\bf r}
-i\omega_{p}\tau} U_{T}(t')\right]
{\cal K}(t',t)U_{T}(t)|\Phi_0(t)\rangle
\Biggr]. 
\end{eqnarray}
The above expression for the {\em total} polarization
contains contributions propagating in various
directions. To obtain the polarization propagating
in a specific direction, one has to expand the
effective--transition operator $U_{T}(t)$ in terms
of $\hat{T}_{1}$ and $\hat{T}_{2}$. Using 
Eqs.\ (\ref{unit})\ and\ (\ref{unit2}) and
keeping only terms contributing to pump/probe
and FWM polarizations, we obtain\cite{per99}
\begin{eqnarray}\label{U}
U^{\dag}_{T}(t) = 
U^{\dag}_{{\rm 1}}(t)
+U^{\dag}_{{\rm 2}}(t) e^{i {\bf k}_{p}{\bf r}}
+ U_{FWM}(t) e^{- 2 i {\bf k}_{p}{\bf r}} +\cdots,
\end{eqnarray}
where (to lowest order in the pump field)
\begin{eqnarray}\label{Upp}
U^{\dag}_1(t)&=&U^{\dag}
+ \frac{1}{2} \left[ \hat{\cal P}(t), \left[U^{\dag},
\hat{\cal P}^{\dag}(t)\right] \right] +
\frac{1}{2} \left[ \hat{\cal P}^{\dag}(t),
\left[ U^{\dag}, \hat{\cal P}(t)\right] \right],
\nonumber\\
U^{\dag}_2(t)&=&
\left[\hat{\cal P}^{\dag}(t), U^{\dag}\right],
\nonumber\\
U_{FWM}^{\dag}(t)&=&
\frac{1}{2}\left[\left[ U,\hat{\cal P}^{\dag}(t)
\right] ,\hat{\cal P}^{\dag}(t)\right]
-\left[U,\hat{\cal P}_2^{\dag}(t)\right]. 
\end{eqnarray}
Here operators $U^{\dag}_1(t)\equiv\tilde{U}(t)$ and
$U_{FWM}^{\dag}(t)$ create one  
{\em e--h} pair, while $U^{\dag}_2(t)$ creates two {\em e--h} pairs
(note that $U_{FWM}$ in Eq.\ (\ref{Upp}) annihilates an {\em e--h} pair).

\subsection*{Pump/probe polarization}

In order to extract the pump/probe polarization
from Eq.\ (\ref{pol}), one should retain only 
terms that are proportional to
$e^{i {\bf k}_{s}\cdot{\bf r}}$. Substituting Eqs.
(\ref{U}) into Eq.\ (\ref{pol}), we obtain
$P_{{\bf k}_{s}}(t)=P^{(1)}_{{\bf k}_{s}}(t)
+P^{(2)}_{{\bf k}_{s}}(t)$,
where 
\begin{eqnarray}\label{PP1}
 P_{{\bf k}_{s}}^{(1)}(t) =-i\mu^2  
e^{i{\bf k}_{s} \cdot {\bf r}- i\omega_{p}
(t - \tau)} \int_{-\infty}^{t} dt' {\cal E}_{s}(t') 
\langle\Phi_0(t)|\tilde{U}(t) {\cal K}(t,t')
\tilde{U}^{\dag}(t') |\Phi_0(t')\rangle,
\end{eqnarray}
and
\begin{eqnarray}\label{PP2}
P_{{\bf k}_{s}}^{(2)}(t) = - i\mu^2   
e^{i{\bf k}_{s} \cdot {\bf r}- i\omega_{p}
(t - \tau)} \int_{-\infty}^{t} dt' {\cal E}_{s}(t')
\langle\Phi_0(t)| U_2(t) {\cal K}(t,t')
U^{\dag}_2(t') |\Phi_0(t')\rangle. 
\end{eqnarray}
Note that the above formulae apply to both undoped
and doped semiconductors. 

Eqs.\ (\ref{PP1})--(\ref{PP2}) express the nonlinear
pump/probe polarization in terms of the linear
response to the probe field of a system described
by the time--dependent effective Hamiltonian
(\ref{eff1}).
Such a form for the nonlinear response allows one to
distinguish between two physically distinct contributions
to the optical nonlinearities.
Assuming that a  short probe pulse arrives at $t=\tau$, consider
the first term, Eq.\ (\ref{PP1}), which gives the
single--pair (exciton for undoped case)  
contribution to the pump/probe polarization. At
negative time delays, $\tau<0$, the probe excites
an {\em e--h} pair, described by state
$\tilde{U}^{\dag}(\tau)|\Phi_0(\tau)\rangle
\simeq U^{\dag}|0\rangle$; since the probe arrives
before the pump, the effective transition operator
coincides with the ``bare''one [see Eqs.\ (\ref{Upp})\
and\ (\ref{sigop})]. The first contribution to the
optical nonlinearities comes from the effective
Hamiltonian, $\tilde{H}(t)$, governing the propagation
of that interacting {\em e--h} pair in the interval
$(\tau, t)$ via the time--evolution operator
${\cal K}(t,\tau)$. Note that since the pump pulse
arrives at $t=0$, for  $|\tau| \gg \Gamma^{-1}$, the
negative time--delay signal vanishes. At $t>0$,  the
{\em e--h} pair (exciton in the undoped case) 
``feels'' the effect of the pump via mainly
the transient bandgap shift, leading, e.g., to
ac--Stark effect, and the change in the band dispersions
(increase in effective mass/density of states), leading to enhanced 
{\em e--h} scattering (exciton binding energy
in the undoped case). Note that  $\tilde{H}(t)$
also contains a  contribution coming from 
the interactions
between probe-- and pump--excited {\em e--h} pairs,
which  are however perturbative in the doped case
for  short pulses or off--resonant excitation
[see Section IV] and  lead to subdominant corrections.
Importantly, 
the response of the system to the optically--induced 
corrections in $\tilde{H}(t)$
takes into account {\em all} orders in the pump field, which is
necessary for the adequate description, e.g., of the
ac--Stark effect and the pump--induced changes in
the {\em e--h} correlations. Indeed, although the pump--induced
term in Eq.\ (\ref{eff1}) is quadratic in ${\cal E}_p$,
the time--evolution of the interacting {\em e--h}
pair is  described without expanding
${\cal K}(t,\tau)$ in the pump field. On the other hand,
the third--order
polarization ($\chi^{(3)}$) can be obtained by expanding 
${\cal K}(t,\tau)$ to the lowest order. The second
contribution to the optical nonlinearities comes from
the matrix element of the final state,
$\langle \Phi_0(t)|\tilde{U}(t)$
in Eq.\ (\ref{PP1}). The latter, given by Eq.\ (\ref{Upp}), 
contains the lowest order (quadratic)
pump--induced terms which describe the Pauli blocking, 
pair--pair, and pair--FS interaction effects (exciton--exciton
interactions in the undoped case\cite{per99}).
 Note that the matrix element of the {\em initial} state
contributes for positive time delays, i.e., if the
probe arrives after the pump pulse. In this case,
however, the pump--induced term in the effective
Hamiltonian (\ref{eff1}) vanishes (since it lasts
only for the duration of the pulse) so that for
positive $\tau> t_p$ the pump/probe signal is
determined by the matrix elements rather than by
$\tilde{H}(t)$. If the probe arrives during the
interaction of the system with the pump pulse 
($\tau\sim t_p$), both  the effective Hamiltonian and
 the matrix elements contribute to the polarization. In
this case, there is also a biexcitonic contribution
[given by Eq.\ (\ref{PP2})] coming from a simultaneous
excitation of two {\em e--h} pairs by the pump
{\em and} the probe. However, such a biexciton state
does {\em not} contribute to negative ($\tau<0$)
time delays. 
As can be seen from the above discussion, 
our theory separates out a number of
contributions that play a different role 
for different 
time delays and excitation conditions.

\subsection*{FWM polarization}

By extracting from Eq.\ (\ref{pol}) all the terms propagating
in the the FWM direction, $2{\bf k}_p-{\bf k}_s$,
we obtain\cite{per99} 
(for delta--function probe ${\cal E}_s(t)={\cal E}_s\delta(t-\tau)$)

\begin{eqnarray}\label{FWM}
 P_{FWM}(t) = - i\mu^2   
e^{i(2{\bf k}_p-{\bf k}_s) \cdot {\bf r}- i\omega_{p}
(t +\tau)} {\cal E}_{s}^{\ast}
&&
\biggl[\langle\Phi_0(t)|U 
{\cal K}(t,\tau)U_{FWM}^{\dag}(\tau)|\Phi_0(\tau)\rangle 
-(t\leftrightarrow \tau)\biggr],
\end{eqnarray}
where the FWM transition operator $U_{FWM}^{\dag}(t)$
is given by Eq.\ (\ref{Upp})
 and $t \ge \tau$. It is convenient to
express $U_{FWM}^{\dag}(t)$ in terms of the ``irreducible''
two--pair operator $W^{\dag}(t)=\frac{1}{2}
\hat{\cal P}^{\dag 2}-\hat{\cal P}_2^{\dag}$, satisfying  
\begin{equation}\label{W}
i\frac{\partial W^{\dag}(t)}{\partial t} = 
\left[\tilde{H}(t),W^{\dag}(t)\right] 
- \mu {\cal E}_p(t)U^{\dag}\hat{\cal P}^{\dag}(t).
\end{equation}
In terms of $W^{\dag}(t)$, the state
$U_{FWM}^{\dag}(t)|\Phi_0(t)\rangle$ in Eq.\
(\ref{FWM}) can be presented as a sum of
two-- and one--pair contributions:
\begin{eqnarray}
\label{Ufwm}
U_{FWM}^{\dag}(t)|\Phi_0(t)\rangle
=UW^{\dag}(t)|\Phi_0(t)\rangle
-\hat{\cal P}^{\dag}(t)U\hat{\cal P}^{\dag}(t)
|\Phi_0(t)\rangle.
\end{eqnarray}
The operator $W^{\dag}(t)$, being quadratic in
the pump field ($\propto {\cal E}_p^2$), describes
the simultaneous excitation of two interacting {\em e--h}
pairs by the pump pulse and includes the two--photon
coherence effects (biexciton and exciton--exciton
scattering effects in the undoped case). The corresponding
state, $W^{\dag}(t)|\Phi_0(t)\rangle$, satisfies
the four--particle (two--exciton for the undoped case)
Schr\"{o}dinger--like equation with a source term
[coming from the second term in the rhs of 
Eq.\ (\ref{W})] and can be expressed in terms of the
two--exciton Green function \cite{per99}. 
 The first term in 
Eq.\ (\ref{Ufwm}) is responsible for the
(interaction--induced) finite FWM signal when the probe
arrives after the pump.\cite{kne98} The second term in 
Eq.\ (\ref{Ufwm}), after being substituted into 
Eq.\ (\ref{FWM}), describes the diffraction of the pump
field on the grating ${\bf k}_p-{\bf k}_s$ due to
the interference of the pump and probe electric fields,
and contributes to the Pauli blocking and
single--exciton effects. Note
that, similarly to the pump/probe polarization, 
Eq.\ (\ref{FWM}) gives contributions to the FWM polarization
in all orders
in the pump field; in this case, however, the third--order
polarization $\chi^{(3)}$  can be obtained by simply
neglecting the difference between  $\tilde{H}(t)$
and $H$ in the time--evolution operator 
${\cal K}(t,\tau)$
and in the equation for $W^{\dag}(t)$
 because $U_{FWM}^{\dag}(t)$
is already quadratic in the pump field. For $\chi^{(3)}$, the
connection between our formalism and that of Axt and
Stahl\cite{axt,kne98} can be established by considering the  matrix
elements of the  operator $W^{\dag}(t)$ between two {\em e--h} pair
states.\cite{per99}

%
%
\section{}
In this appendix we clarify our convention for
the time delay $\tau$ and relate it to the most
commonly used conventions in Pump/probe
and Four Wave Mixing experiments. In the generic
experimental configuration two laser pulses
$E_{1}(t)e^{i{\bf k}_1 \cdot {\bf r} - i\omega (t-t_1)}$,
and
$E_{2}(t)e^{i{\bf k}_2 \cdot {\bf r} - i\omega (t-t_2)}$,
respectively centered at time $t= t_1$ and $t= t_2$ are
incident on a sample. Let us define $\Delta t$ as,
\begin{equation}
\Delta t =  t_1 - t_2
\end{equation}
and consider a FWM experiment where the signal is
measured in the direction
$2{\bf k}_2 - {\bf k}_1$. Then
for a two-level-atom, the signal vanishes for
$\Delta t <0$, while  for $\Delta t>0$  its amplitude,
which decays with time as $e^{-t/T_2}$, is determined by 
the Pauli blocking. In a system with Coulomb interactions 
(such as  a semiconductor) a FWM signal is observed both
for $\Delta t<0$ and $\Delta t>0$. The $\Delta t <0$
signal is entirely due to the Coulomb interaction.

In pump/probe experiments, one usually chooses 
one pulse (the ``pump'') to have an amplitude $E_{p}$
much larger than that of the other pulse (the
``probe''),  $E_{s}$. As discussed in the text, a weak 
probe measures the linear
response of the system (bare or dressed by the
pump). If we choose that $E_{p} = E_{2}$,  the pump
induces coherent and incoherent populations when
it arrives in the sample {\em before} the probe
i.e. for $t_2<t_1$. This is usually defined as
``positive'' time delay $\tau =  t_2 - t_1 >0$ in
the pump/probe literature. Note that
$\tau =  - \Delta t$, i.e., the ``regular'' sequence
in pump/probe experiments is the {\em reverse} of
that of FWM experiments. For $\tau <0$, the origin
of the pump/probe signal is that the probe creates
a linear polarization in the sample which lasts for
time $\sim T_2=\Gamma^{-1}$ and, consequently, is
scattered by polarization 
excited by the pump field. The signal
observed for $\tau <0$ is therefore due to coherent
effects.

In FWM experiments,
there is no restriction on the magnitude of the two
incident fields $E_{2}$ and $E_{1}$, which are often
chosen to have amplitudes of the same order.
 Note however that, at the $\chi^{(3)}$ level, the FWM
and pump--probe polarizations are linear
in the $E_{1}(t)$ field
and thus the above linear response calculation 
applies even for comparable pump and probe amplitudes.

%
%
\section{}

In this Appendix  we present the explicit expressions 
for the renormalized transition matrix elements in the presence of
the pump excitation. 
The direct transition matrix element is given by
\begin{eqnarray}
M_{{\bf p}}(t) &&= 1 - \left| {\cal P}_{eh}({\bf p},t)\right|^{2} 
+ \Biggl[ {\cal P}^{*}_{{\rm eh}}({\bf p},t) 
\sum_{k'<k_{F}} {\cal P}_{eh}^{{\rm e}}({\bf pk';k'};t)
+ {\rm H.c.} \Biggr] 
\nonumber\\&&
-\frac{1}{2} \sum_{p'>k_{F}} {\cal P}_{eh}^{*}({\bf p'},t) 
\Biggl[ {\cal P}_{eh}^{{\rm e}}({\bf p' p; p'};t)
-{\cal P}_{eh}^{{\rm h}}({\bf p' p;p'}; t) 
-{\cal P}_{eh}^{{\rm h}}({\bf p' p ; p};t)   
+  {\cal P}_{eh}^{{\rm e}}({\bf p' p; p};t) \Biggr] 
\nonumber\\&& 
+ \frac{1}{2} \sum_{p'>k_{F}}
\Biggl[{\cal P}_{eh}({\bf p},t)  + 
{\cal P}_{eh}({\bf p'},t)\Biggr] 
\Biggl[ {\cal P}_{eh}^{{\rm e} }({\bf p p'; p};t)
+ {\cal P}_{eh}^{{\rm h}}({\bf p p'; p};t)\Biggr]^{*} \label{me1}.
\end{eqnarray}
The first term on the rhs of the above equation describes the phase
space filling contribution, 
while the rest of the terms are due to  the 
mean field  pair--pair and pair--FS interactions. 

The pump--induced indirect transition matrix element is given by
\begin{eqnarray}
M_{{\bf p p' k}}(t) = 
&&
\Biggl[ {\cal P}_{eh}({\bf k},t) - {\cal P}_{eh}
({\bf p + p' - k},t) \Biggr]^{\ast}  
{\cal P}_{eh}^{{\rm e}} ({\bf p  p';k};t)
\nonumber\\&&
-\Biggl[{\cal P}_{eh}({\bf p},t)+{\cal P}_{eh}({\bf p'},t)\Biggr]^{\ast}  
\Biggl[{\cal P}_{eh}^{{\rm e}} ({\bf p  p';p};t) 
+ {\cal P}_{eh}^{{\rm h} *}({\bf p  p';p+p'-k};t)\Biggr] 
\nonumber\\&&
+ {\cal P}_{eh}({\bf p+p'-k},t) 
\Biggl[ {\cal P}_{eh}^{{\rm e}}({\bf k,p+p'-k;p'};t) 
- {\cal P}_{eh}^{{\rm e} }({\bf k,p+p'-k;p};t)\Biggr]^{\ast}  
\nonumber\\&&
+ {\cal P}_{eh}^{\ast}({\bf k},t) 
\Biggl[ {\cal P}_{eh}^{{\rm h}}({\bf k,p+p'-k;p'};t)
+ {\cal P}_{eh}^{{\rm e}}({\bf p ,p';k};t)
\nonumber\\&&
- {\cal P}_{eh}^{{\rm e}}({\bf p , p'; p+p'-k};t)
-{\cal P}_{eh}^{{\rm h}}({\bf k,p+p'-k;p};t) \Biggr]
\nonumber\\&&
+ {\cal P}_{eh}({\bf p},t)  {\cal P}_{eh}^{{\rm e}*}({\bf k, p+p'-k; p'};t)
- {\cal P}_{eh}({\bf p'},t)  {\cal P}_{eh}^{e *}({\bf k, p+p'-k;p};t) 
\label{me2}.
\end{eqnarray} 

The effective {\em e--h} potential is given by
\begin{eqnarray}\label{veh}
\upsilon_{eh}({\bf q; kk'};t)= 
&&
\upsilon({\bf q})  -  \frac{\mu}{2}\,{\cal E}_p(t) 
\Biggl[ {\cal P}_{eh}^{e}({\bf k+q,k';k'+q};t) 
+{\cal P}_{eh}^{h}({\bf k , k'+q;k'};t)  \Biggr] 
\nonumber\\&&
- \frac{\mu}{2}\,{\cal E}_p(t) 
\Biggl[ {\cal P}_{eh}^{e}({\bf k,k'+q;k'};t) + 
{\cal P}_{eh}^{h}({\bf k+q, k';k'+q};t)
\Biggr]^{\ast},
\end{eqnarray} 
and the effective {\em e--e} potential is given by
\begin{eqnarray} \label{vee}
\upsilon_{ee}({\bf q;k k'};t) =  
&&
\upsilon({\bf q})  + \frac{\mu}{4}\, {\cal E}_p(t)
\Biggl[ {\cal P}_{eh}^{e}({\bf k+q,k'-q;k'};t) - 
{\cal P}_{eh}^{e}({\bf k+q,k'-q;k};t)\Biggr] 
\nonumber\\&&
+ \frac{\mu}{4}\, {\cal E}_p(t)
\Biggl[{\cal P}_{eh}^{e}({\bf k,k';k'-q};t) 
-{\cal P}_{eh}^{e}({\bf k,k';k+q;}t) \Biggr]^{\ast}.
\end{eqnarray} 
%



\begin{figure}
\caption{
The effect of the pump pulse on the band dispersion,
$\varepsilon_{{\bf k}}(t) 
=\varepsilon^{c}_{{\bf k}}(t)  + \varepsilon^{v}_{-{\bf k}}(t)$, 
of the ``pump--dressed'' system. 
Solid line: ``bare'' dispersion ($t=-\infty$). Dashed line:
pump--renormalized dispersion ($t=0$). The bands become ``heavier''
for the duration of the pump.
}
\end{figure}

\begin{figure}
\caption{
(a) The {\em e--h} scattering processes that contribute to the rhs of
Eq.\ (\ref{de}). Full lines correspond to $s({\bf p,k},t)$ and thin lines
to $V$. The diagrams describe (from top to bottom) (i) Born
scattering of a FS electron,  
(ii) FS electron ladder diagrams,  
(iii) FS hole ladder diagrams, and 
(iv) Nonlinear vertex corrections due to the dynamical FS response.
(b) Scattering processes state that
determine the time-- and momentum--dependence 
of the  effective {\em e--h} potential  
$\tilde{V}({\bf p},t) $ and 
lead to the unbinding of the HFA bound
state.}
\end{figure}

\begin{figure}
\caption{
Linear absorption resonance lineshape 
for FES (solid curve) compared to the HFA (dashed curve), 
calculated with $g=0.4$ and $\Gamma=0.1 E_{F}$. 
The HFA resonance position was shifted for better visibility.}
\end{figure}

\begin{figure}
\caption{The absorption spectra for the FES (a) compared 
to the HFA (b). Solid curves: linear absorption. 
Dashed curves: nonlinear  absorption.
The curves were calculated with 
$g=0.4$ and $\Gamma =0.1 E_{F}$ at time 
delay $\tau=-0.1\Gamma^{-1}=-t_{p}/2$ 
and pulse duration $t_{p} = 2.0E_{F}^{-1}$.
The nonlinear absorption lineshapes exhibit bleaching, 
resonance blueshift, and gain below
the absorption onset
that differ in the two cases.}
\end{figure}

\begin{figure}
\caption{
Differential transmission lineshape for various time delays calculated with 
$g=0.4$, $\Gamma =0.1 E_{F}$, and $t_{p} = 2.0E_{F}^{-1}$.
(a) Long time delay, $\tau = - 1.5\Gamma^{-1}=-15.0 E_F^{-1}$. 
For the FES, the oscillations in
the differential transmission spectra are reduced (solid curve)
as compared to the HFA (dashed curve). 
(b) Short time delay, 
$\tau = - 0.1 T_2 = - t_p/2 = - 1.0 E_F^{-1}$.
For the FES, the differential transmission spectrum 
is asymmetric (solid curve) as compared to the
symmetric lineshape for the HFA (dashed curve).
The above curves were shifted for better visibility.
}
\end{figure}

\begin{figure}
\caption{
(a) Nonlinear absorption resonance bleaching 
evaluated at the instantaneous peak frequency 
as function of time delay 
for the FES (solid curve) compared to the HFA (dashed curve). The
curves were calculated with 
$g=0.4$, $\Gamma =0.1 E_{F}$, and $t_{p} = 2.0E_{F}^{-1}$.
The time--dependence of the pump pulse 
is also presented for comparison (dotted curve).
(b) Same calculated using the rigid band shift model, Eq.\ (\ref{nla}).
}
\end{figure}

\begin{figure}
\caption{
The enhancement of the nonlinear
absorption resonance of the  FES (dashed curve) vs. linear
absorption (solid curve) at long time delay $\tau=-1.5 \Gamma^{-1}$.
The curves were calculated with 
$g=0.4$, $\Gamma =0.1 E_{F}$, and $t_{p} = 2.0E_{F}^{-1}$.
}
\end{figure}

\begin{figure}
\caption{
Nonlinear absorption resonance blueshift as function of time delay 
for the FES (solid curve) compared to the HFA (dashed curve).
The blueshift is significantly weaker for the HFA.
The curves were calculated with 
$g=0.4$, $\Gamma =0.1 E_{F}$, and $t_{p} = 2.0E_{F}^{-1}$.
The time--dependence of the pump pulse 
is also presented for comparison (dotted curve). 
}
\end{figure}

\begin{figure}
\caption{
Resonance bleaching as function of time delay 
for the FES (a) compared to the HFA (b) for 
different strengths of the {\em e--h} interaction: 
g=0.4 (solid curve) and g=0.3 (dashed curve).
The curves were  calculated with
$\Gamma =0.1 E_{F}$, and $t_{p} = 2.0E_{F}^{-1}$.
The time--dependence of the pump pulse 
is also presented for comparison (dotted curve). 
}
\end{figure}

\begin{figure}
\caption{The effect of the pump--induced renormalization of the band
dispersions on the {\em e--h} interactions. The 
function $F(\omega,\tau)$, given by Eq.\ (\ref{F}), for the
FES (a) compared to the HFA (b) in the presence (dashed curve)
and absence (solid curve) of the pump pulse.
The curves were calculated with 
$g=0.4$, $\Gamma =0.1 E_{F}$, and $t_{p} = 2.0E_{F}^{-1}$.
}
\end{figure}

\begin{figure}
\caption{
The effect of the pump--induced 
renormalization of the band dispersions on the 
resonance strength.
The nonlinear absorption spectrum for the 
FES (a) compared to the HFA (b). 
Solid curves: Linear absorption. 
Dashed curves: Nonlinear absorption. 
Dotted curves: 
Nonlinear absorption for a rigid band shift only. 
The curves were  calculated with 
$g=0.4$, $\Gamma =0.1 E_{F}$, and $t_{p} = 2.0E_{F}^{-1}$.
}
\end{figure}

\clearpage

\epsfxsize=6.0in
\epsffile{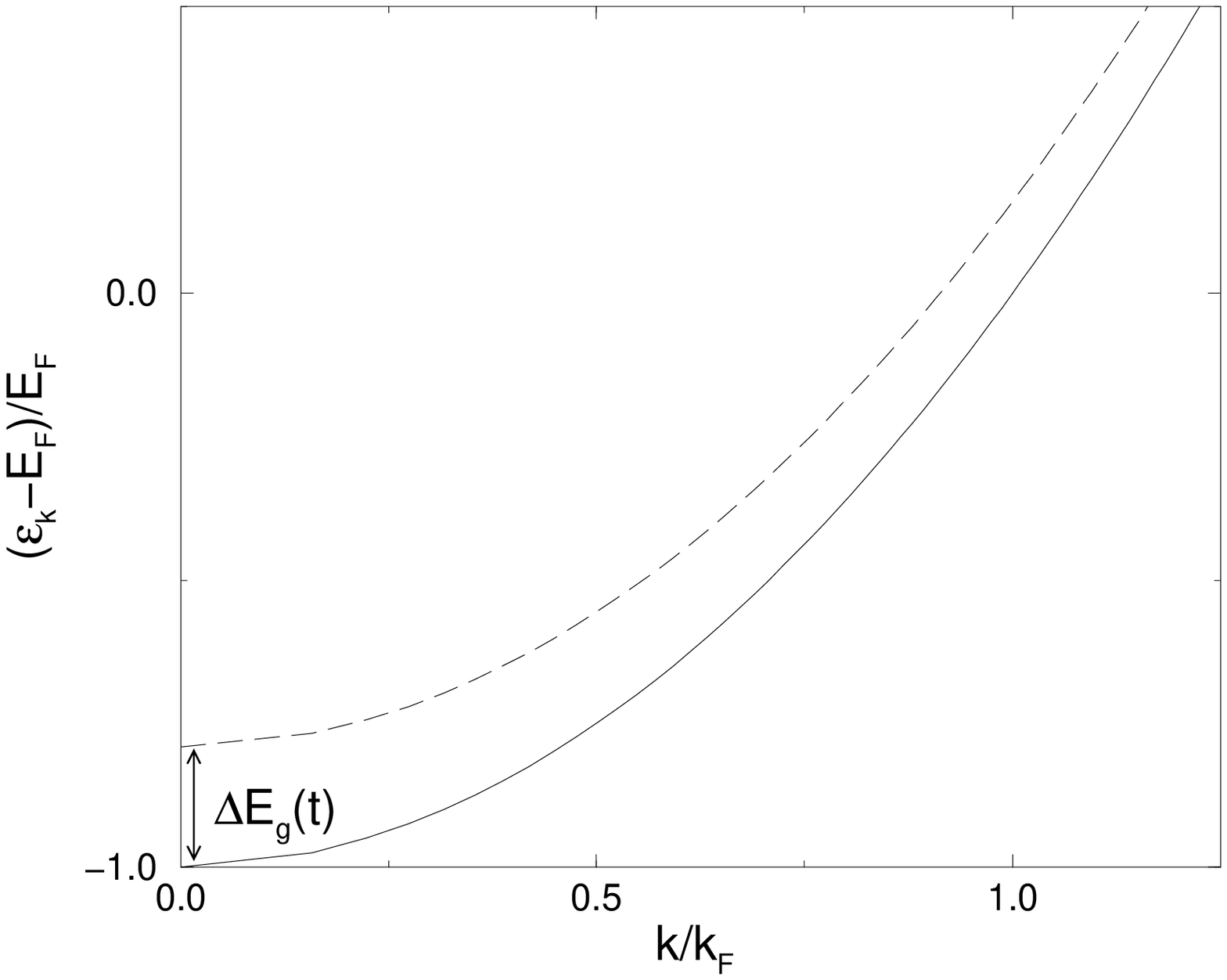}
\vspace{40mm}
\centerline{FIG. 1}
\epsfxsize=4.0in
\epsffile{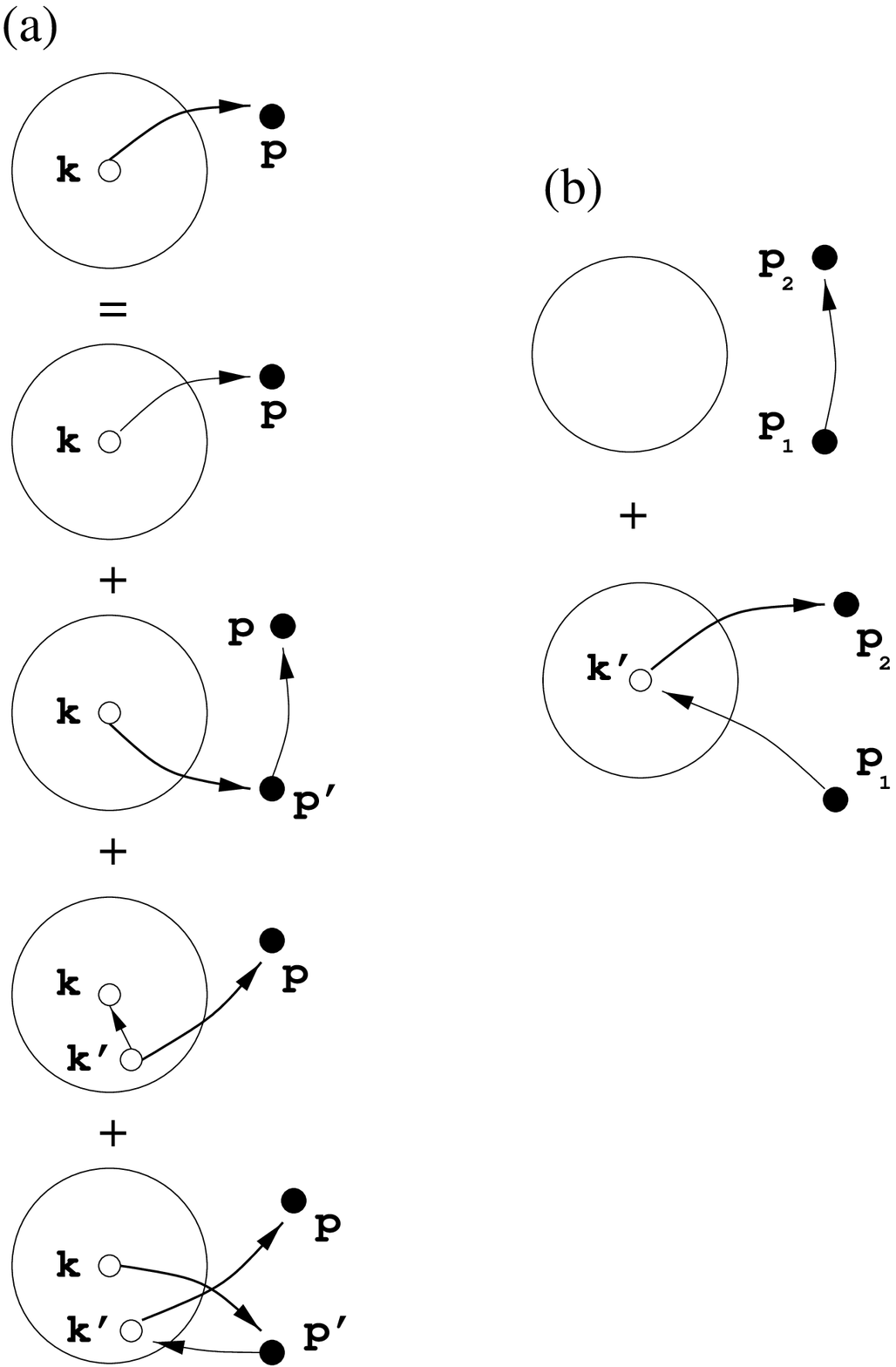}
\vspace{40mm}
\centerline{FIG. 2}
\epsfxsize=6.0in
\epsffile{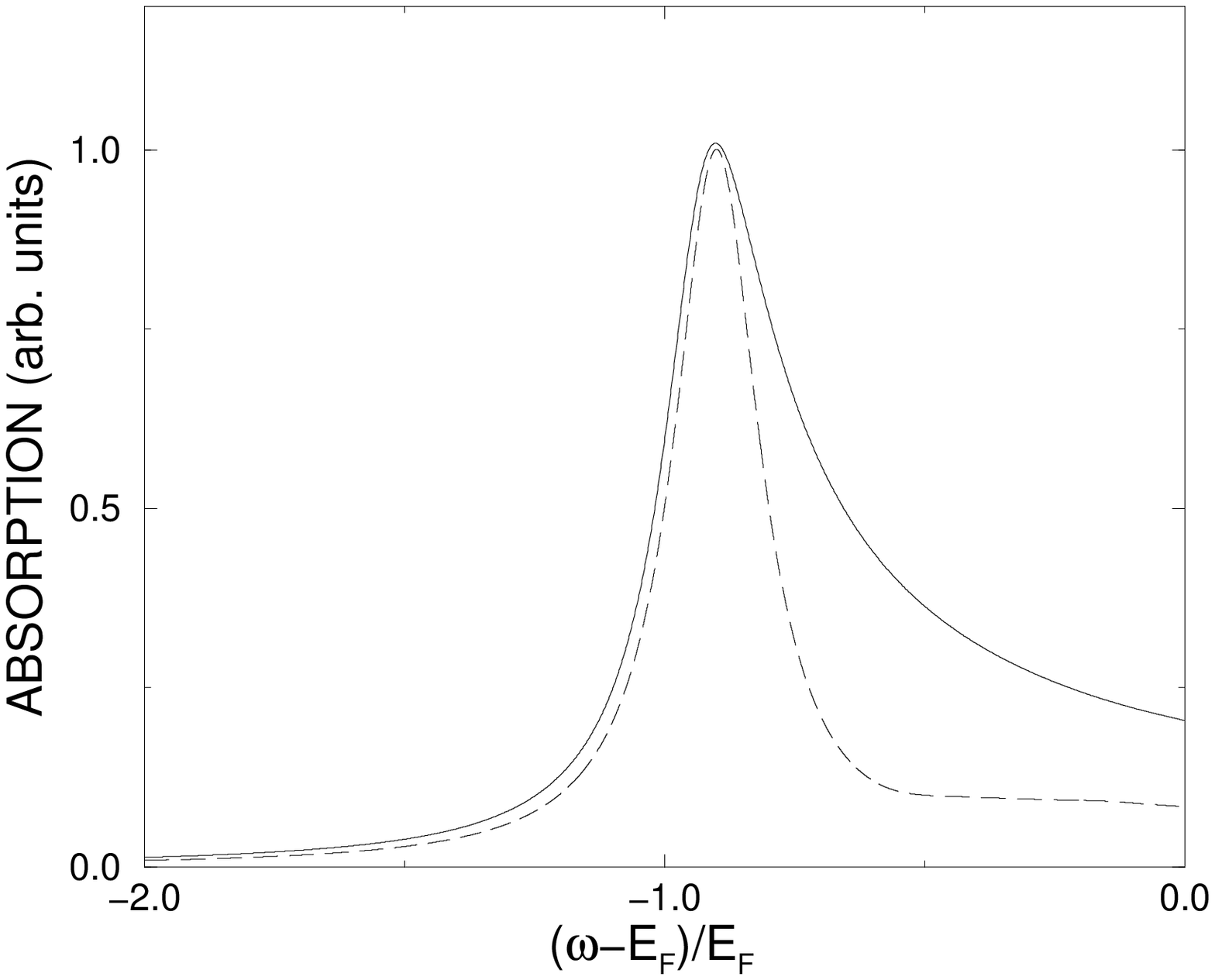}
\vspace{40mm}
\centerline{FIG. 3}
\epsfxsize=6.0in
\epsffile{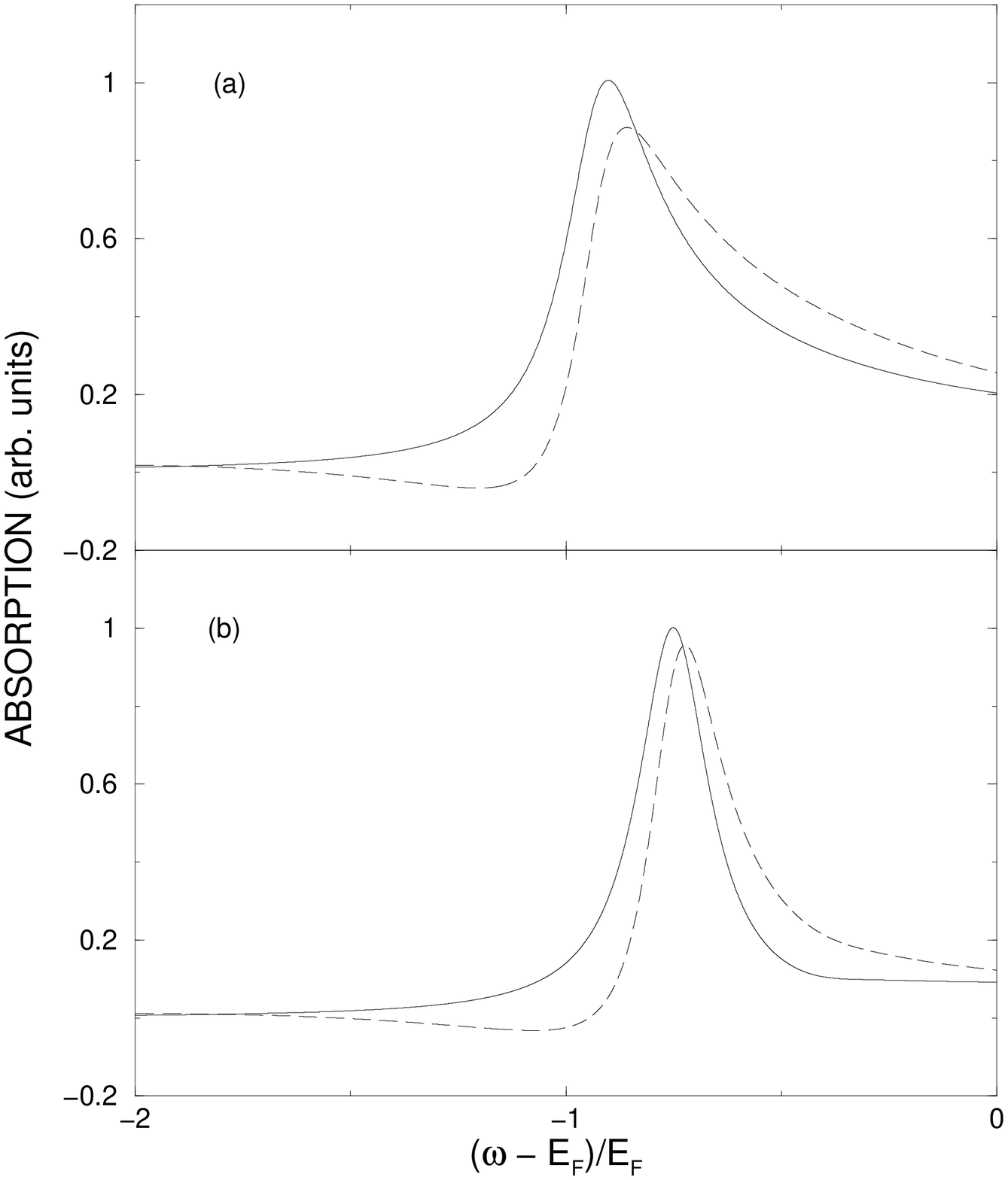}
\vspace{20mm}
\centerline{FIG. 4}
\epsfxsize=6.0in
\epsffile{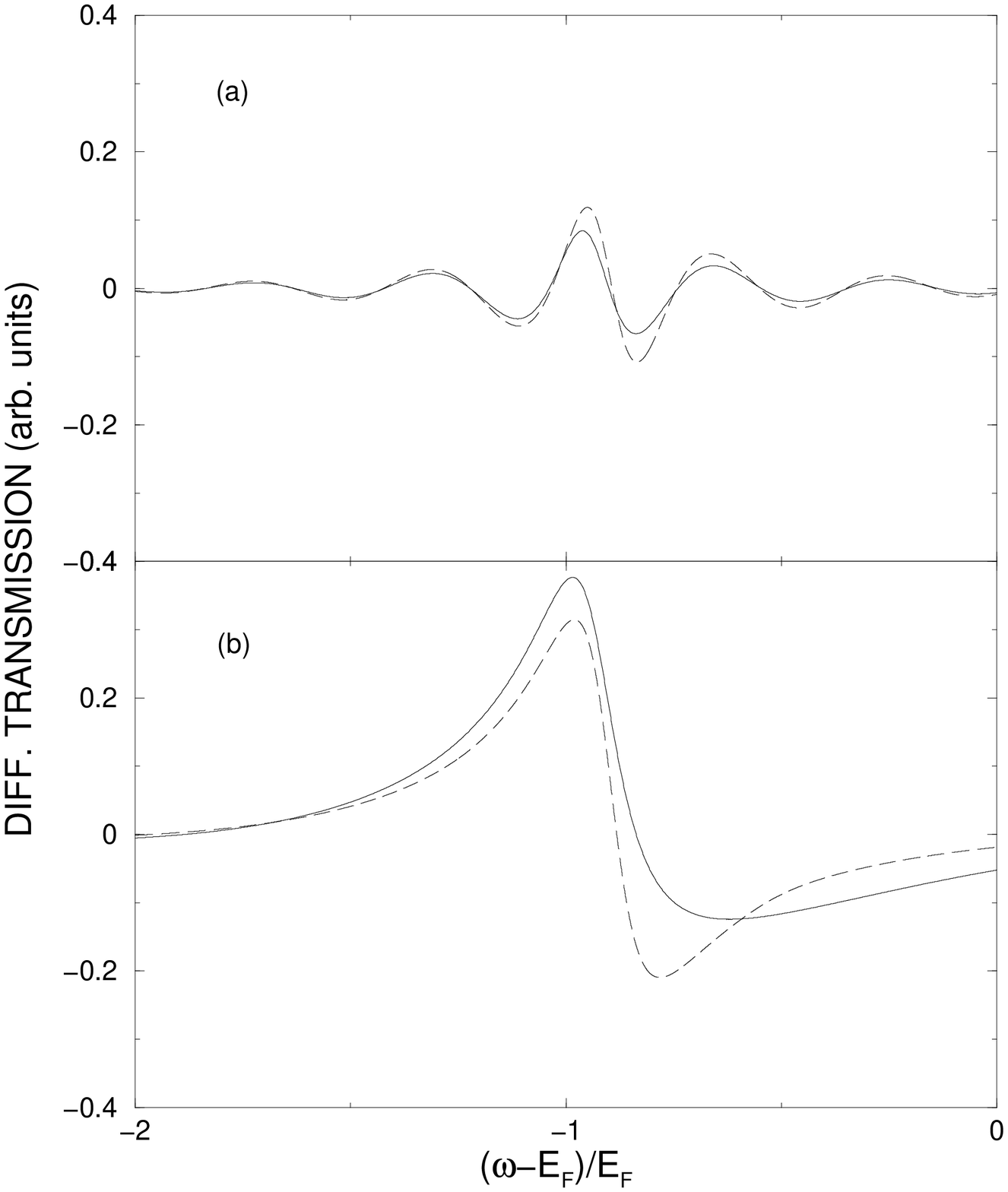}
\vspace{20mm}
\centerline{FIG. 5}
\epsfxsize=6.0in
\epsffile{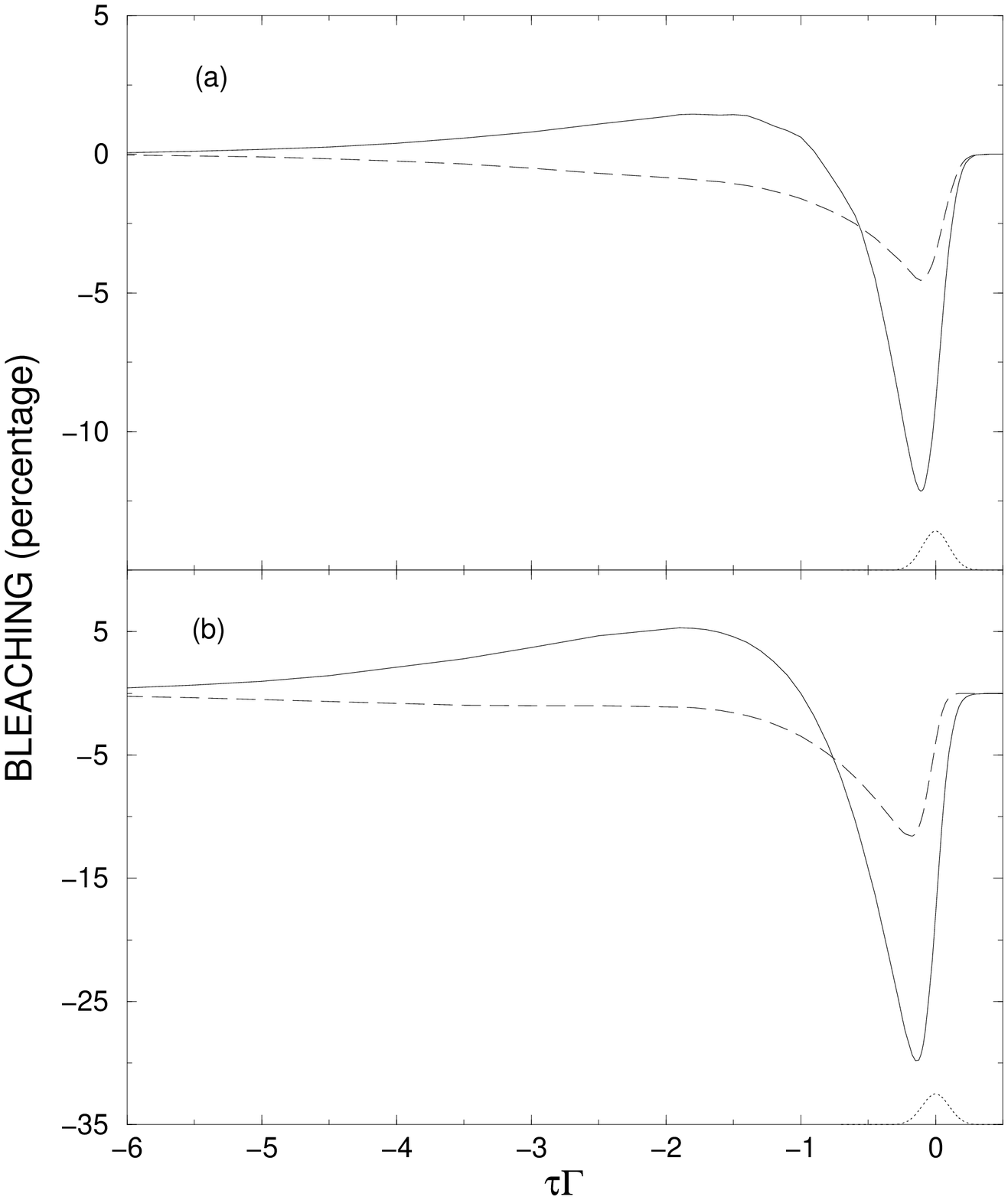}
\vspace{20mm}
\centerline{FIG. 6}
\epsfxsize=6.0in
\epsffile{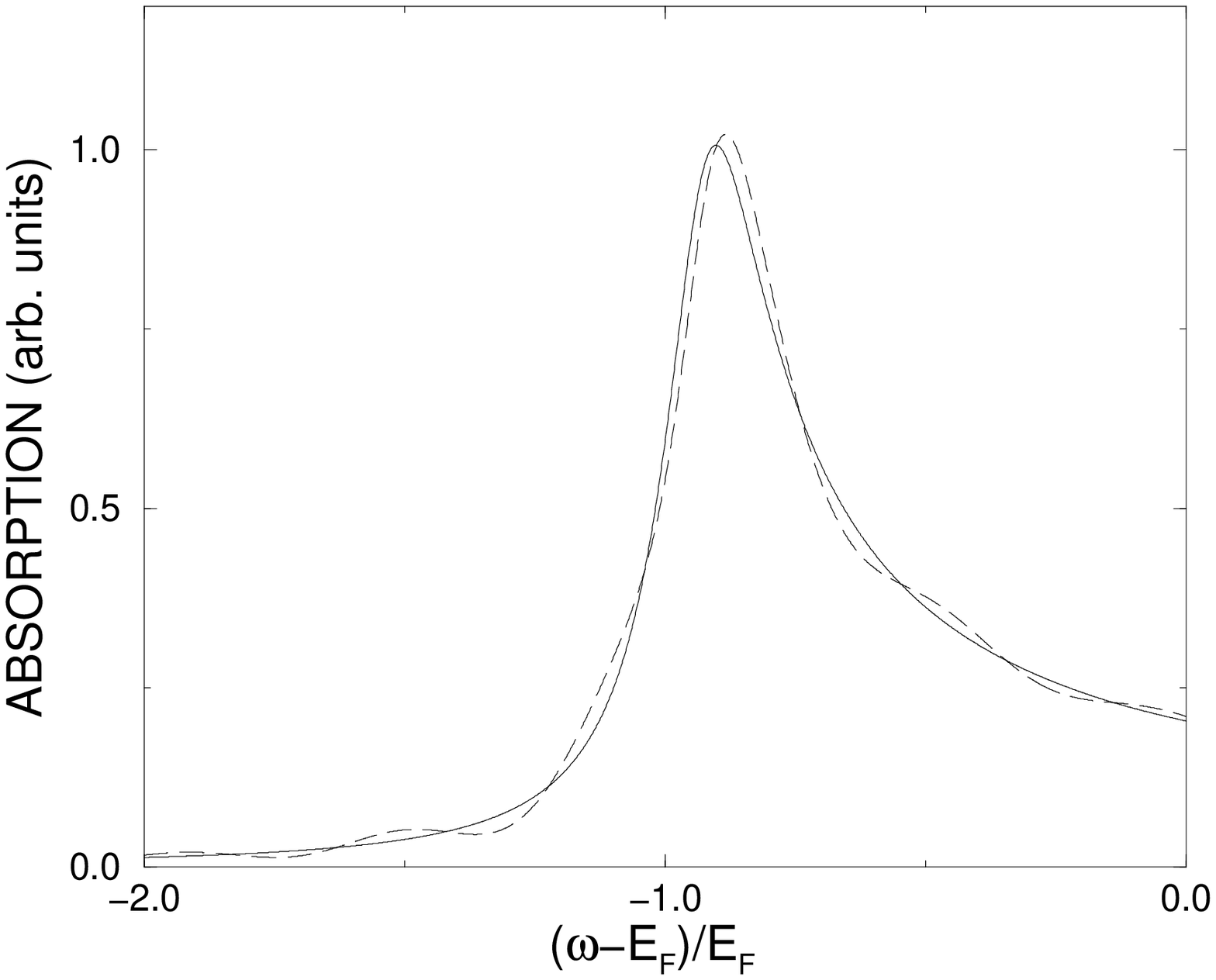}
\vspace{40mm}
\centerline{FIG. 7}
\epsfxsize=6.0in
\epsffile{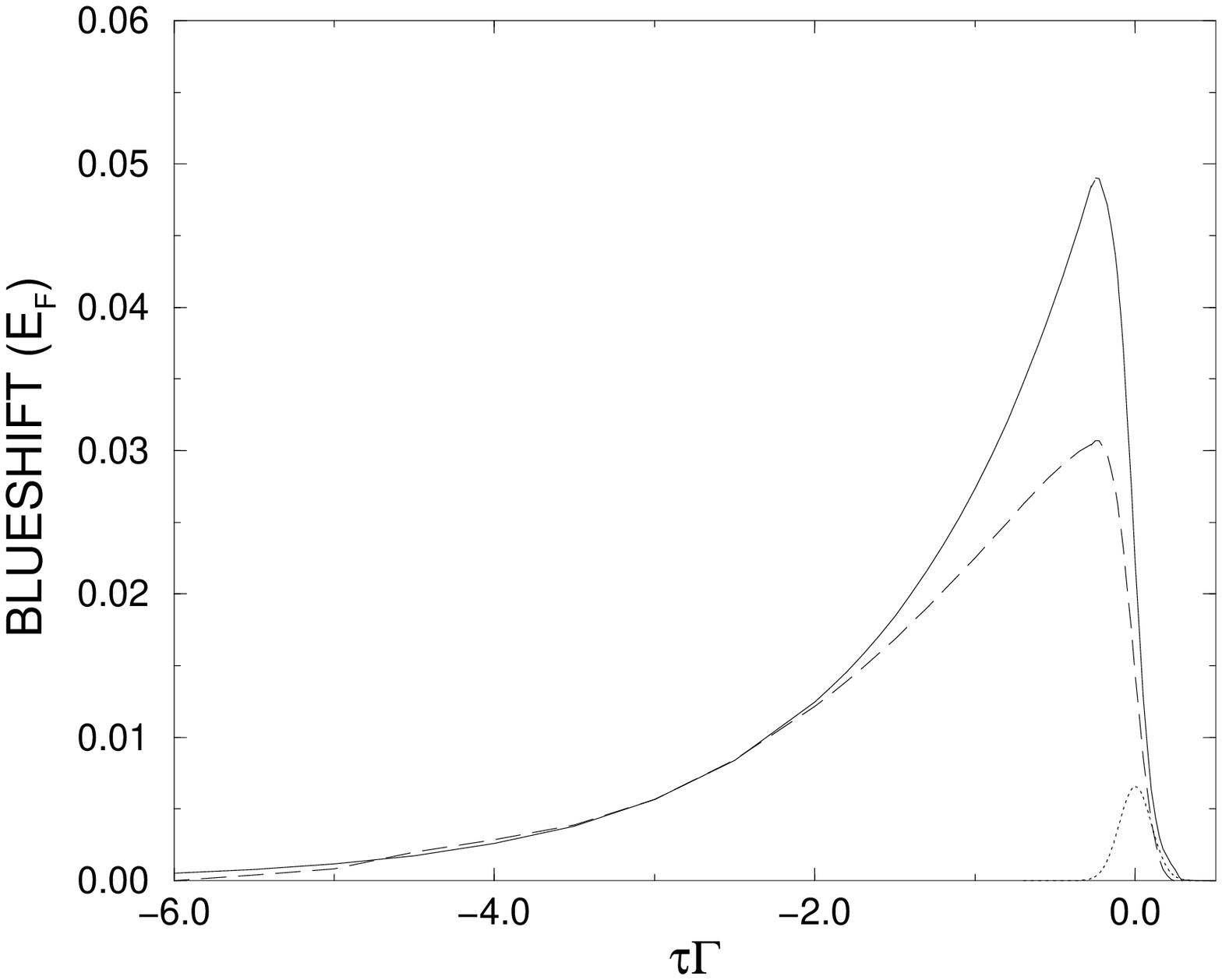}
\vspace{40mm}
\centerline{FIG. 8}
\epsfxsize=6.0in
\epsffile{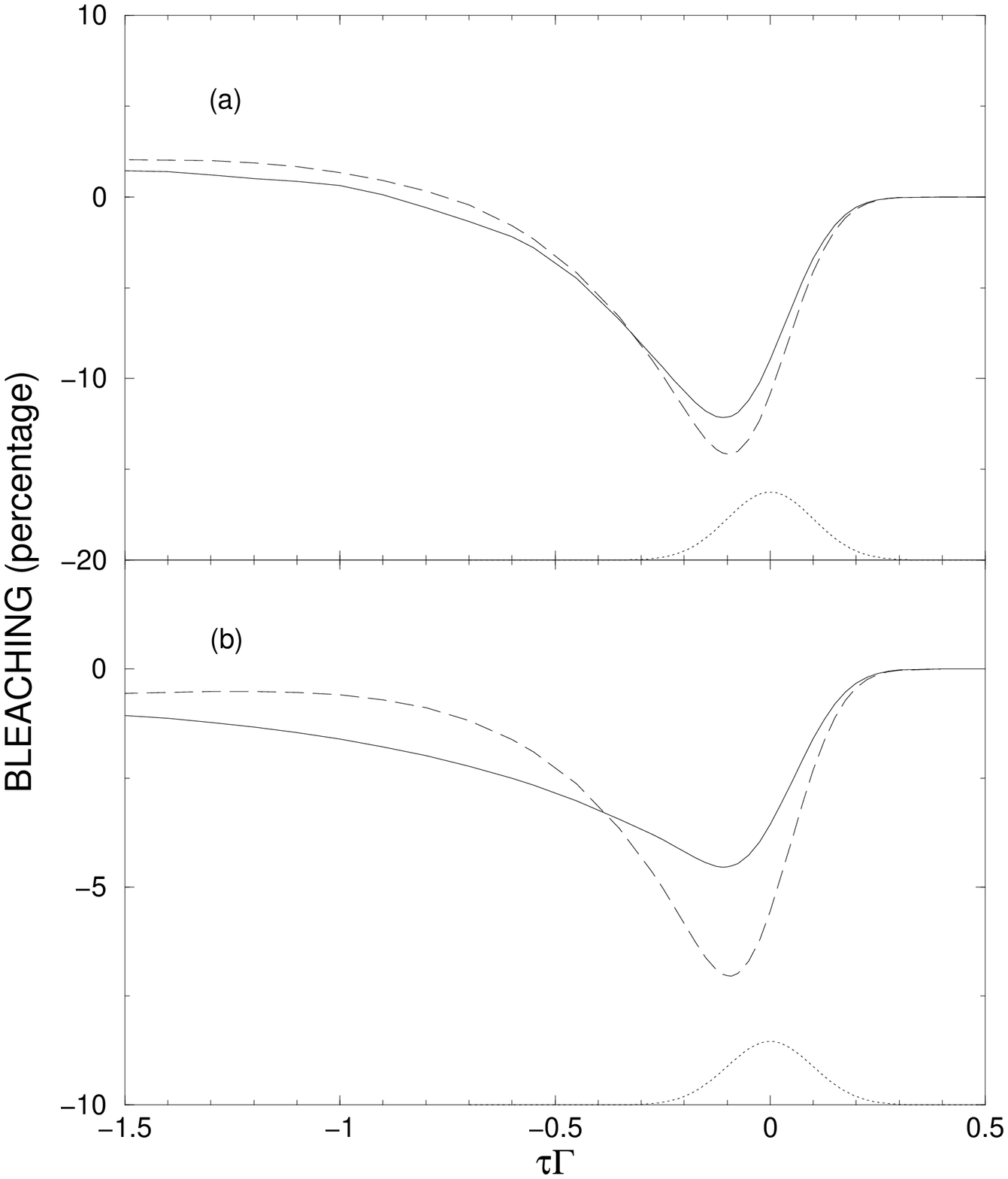}
\vspace{20mm}
\centerline{FIG. 9}
\epsfxsize=6.0in
\epsffile{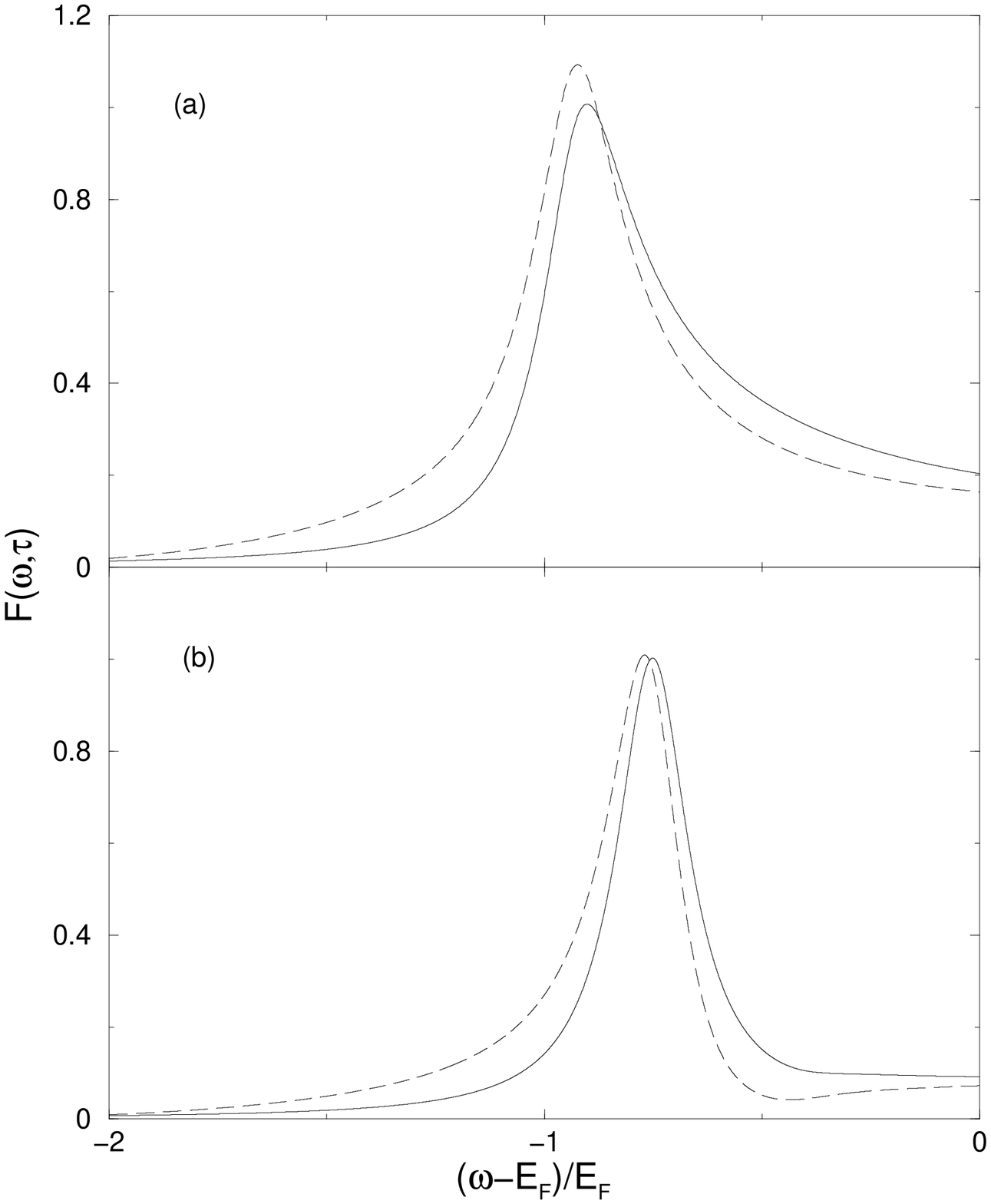}
\vspace{20mm}
\centerline{FIG. 10}
\epsfxsize=6.0in
\epsffile{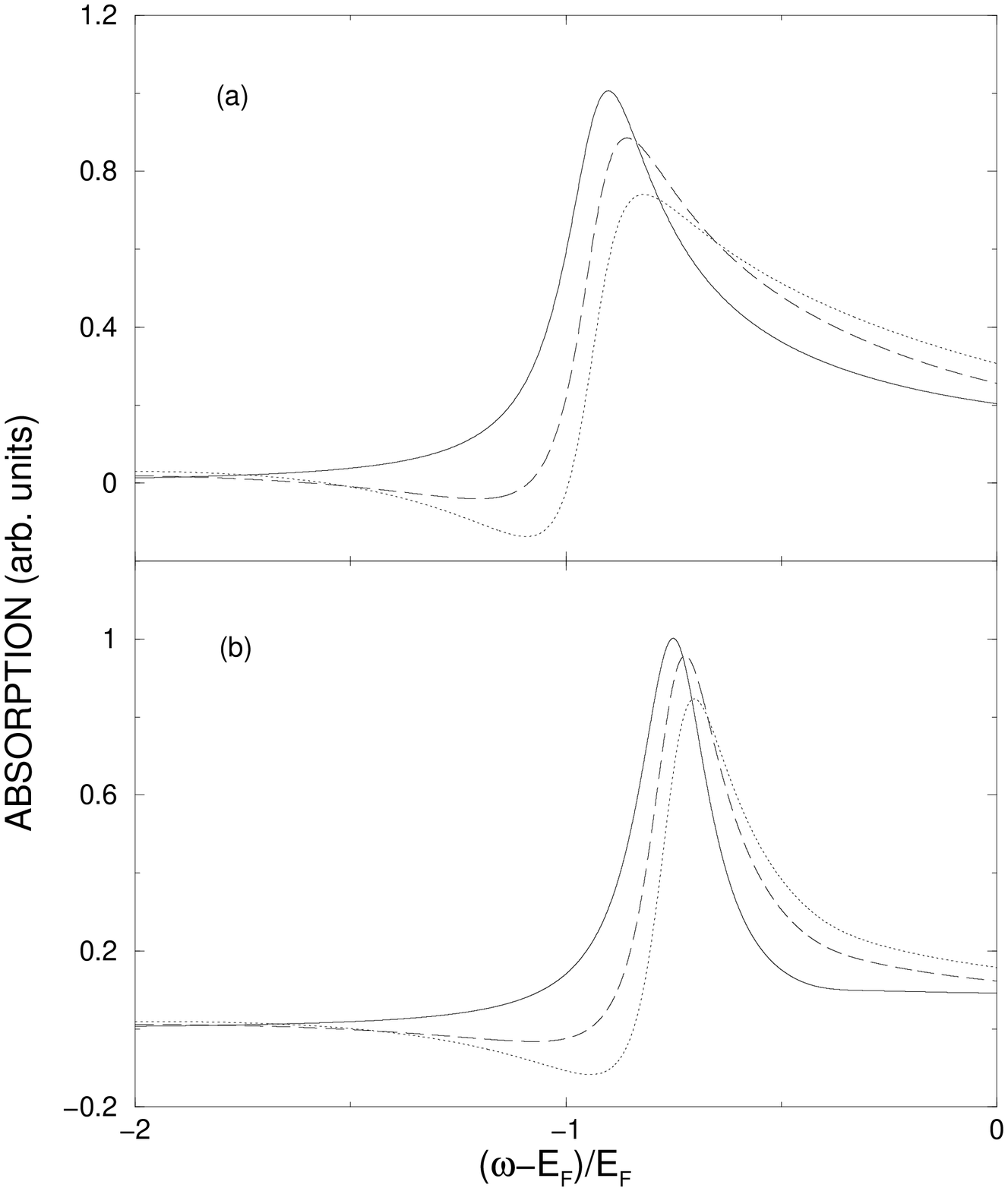}
\vspace{20mm}
\centerline{FIG. 11}

\end{document}